\documentstyle[12pt]{article}

\newcommand{\tr}{{\rm tr}}
\newcommand{\Tr}{{\rm Tr}}

\title{  
The Gauged Thirring Model 
} 
\author{%
  Kei-Ichi Kondo
  \thanks{E-mail: kondo@cuphd.nd.chiba-u.ac.jp}
  \vspace{1.2em}  
  \\
  {\it HLRZ c/o Forschungszentrum, KFA J\"ulich}\\ 
  {\it D-52425 J\"ulich, Germany}
  \thanks{Address from September 1995  to February 1996.
  On leave of absence from: 
  Department of Physics, Faculty of Science,
  Chiba University, Chiba 263, Japan.
  Address from March 1996  to December 1996:
  Department of Physics, Theoretical Physics, 
  University of Oxford,
  1 Keble Road, Oxford, OX1 3NP, UK.}
}
\date{CHIBA-EP-93/HLRZ-1/96, \\
January 1996, \\
hep-th/9603151}
\begin{document}
\maketitle
%\newpage
\centerline{Abstract} 
We propose the gauged Thirring model as a natural
gauge-invariant generalization of the Thirring model,
four-fermion interaction of current-current type.
In the strong gauge-coupling limit, the gauged Thirring
model reduces to the recently proposed reformulation of the
Thirring model as a gauge theory.
Especially, we pay attention to the effect coming from the
kinetic term for the gauge boson field, which was
originally the auxiliary field without the kinetic term.
In 3 + 1 dimensions, we find the nontrivial phase
structure for the gauged Thirring model, based on
the Schwinger-Dyson equation for the fermion propagator as
well as the gauge-invariant effective potential for the
chiral  order parameter.  
Within this approximation, we study the renormalization
group flows (lines of constant physics) and find
a signal for nontrivial continuum limit with nonvanishing
renormalized coupling constant and large anomalous dimension
for the gauged Thirring model in 3+1 dimensions, at least
for small number of flavors $N_f$. Finally we discuss the
(perturbatively) renormalizable extension of the gauged
Thirring model.

\newpage
\section{Introduction}
\setcounter{equation}{0}

The Nambu--Jona-Lasinio (NJL) model \cite{NJL} of
four-fermion interaction is a successful model to
study the chiral dynamics.  
Actually, it has succeeded to extract many characteristic
features of the chiral dynamics.  However, it is usually
thought that the NJL model lacks the sound basis as a
quantum field theory, namely, it is (perturbatively)
nonrenormalizable.  Therefore, the NJL model is regarded as
an effective quantum field theory
which is nevertheless quite important phenomenologically. 
\par
Indeed, the NJL model in 3+1 dimensions is  
non-renormalizable in the usual perturbation theory in the
coupling constant. Nevertheless, it has been established
that a class of the 2+1 dimensional four-fermion model,
e.g. the Gross-Neveu model, is renormalizable in the
different expansion scheme,
$1/N_f$ expansion \cite{RWP91}. However, there is no
evidence that the 3+1 dimensional NJL model becomes
renormalizable by choosing some clever expansion scheme.
Moreover, it has been shown on the lattice that the 3+1
dimensional NJL model is not renormalizable even in the
non-perturbative sense
\cite{Ali95}.
\par
It is generally believed that 
the non-perturbatively as well as
perturbatively non-renormalizable model  leads to trivial
(non-interacting free) theory in the continuum limit.  
Here the continuum limit means the procedure of obtaining
the well-defined finite continuum field theory  from the
regularized model by taking the limit of removing the
cutoffs, e.g. taking the $a \rightarrow 0$ ($a$: lattice
spacing) limit in the lattice field theory and taking the
$\Lambda \rightarrow \infty$ limit in the model with an
ultraviolet cutoff $\Lambda$, etc. 
Hence, if a theory is nontrivial, it is expected that
the theory can be renormalizable in a certain
renormalization scheme.  The authors of ref.~\cite{HKKN94}
use the following terminology: they call a theory 
(non-perturbatively) {\it renormalizable} if the theory is
perturbatively non-renormalizable and nevertheless gives a
nontrivial continuum limit, rather than simply calling it
nontrivial. This is indeed consistent with the
picture of modern renormalization group (RG) in Wilson's
sense \cite{Wilson74}, namely, the nontrivial continuum
limit is realized on a nontrivial fixed point if the
cutoff theory approaches to the fixed point in the bare
parameter space along the renormalized trajectory as a
result of having performed the successive renormalization
steps.
\par
\par
Recently, another possibility of renormalizability (in
the sense defined above) for a class of models with
four-fermion interaction was suggested in
\cite{KSY91}. 
The essential ingredient is the introduction
of the gauge interaction into the NJL model.  If the
gauge interaction is incorporated, the model is called the
gauged NJL model
\cite{BLL86}.  The gauged NJL model was initiated by
Bardeen et al.
\cite{BLL86} motivated from the renormalization group
consideration of the strongly coupled gauge theory in the
sense of Miransky
\cite{Miransky85,Miransky93}. It has been found that the
gauged NJL model has nontrivial phase structures
\cite{KMY89,BLL89}. In the work \cite{KSY91}, it was
claimed that  the gauge interaction might promote the
trivial NJL model to an
 (non-perturbatively renormalizable) interacting theory. 
It was suggested in \cite{KSY91,Yamawaki92} that the
gauged NJL model is renormalizable if and only if the
equivalent gauge-Higgs-Yukawa system approaches the
infrared fixed point of Pendleton and Ross
\cite{PR81} (PR) in the continuum limit and that, for the
renormalizability of the gauged NJL model, the presence of
asymptotically free (or fixed coupling as a standing limit
of slowly running \cite{KMY89}) gauge interaction is
essential, but the asymptotic freedom of the gauge
interaction should not be too strong. Such observations
were also claimed by Krasnikov
\cite{Krasnikov93} in the study of the gauged NJL model in
$4-\epsilon$ dimensions. 
This situation is in sharp contrast with the Abelian-gauged
NJL model with the non-asymptotic free (running) gauge
coupling.  In such a class of the gauged NJL model, there is
no indication of the nontrivial fixed point on which the
nontrivial continuum may be obtained
\cite{Kondo91a,Kondo91b,Rakow91}.
\par
Actually, the renormalizability of the gauged NJL model in
3+1 dimensions has been shown within the ladder
approximation of the Schwinger-Dyson (SD) equation
\cite{KTY93}.  And this conclusion has further been
supported by the subsequent work \cite{KSTY94}
in which the flow of the renormalized Yukawa coupling
was analyzed in the corresponding gauge-Higgs-Yukawa
theory.
These works are, however, restricted to the fixed
coupling case in the ladder approximation of the SD
equation.  
In the quite recent work of Kyoto group lead by
Kugo \cite{HKKN94}, it was proved in the scheme of $1/N_c$
expansion that the gauged NJL model (as a limit of a
gauge-Higgs-Yukawa system which is perturbatively
renormalizable) is renormalizable in 3+1 dimensions in the
leading order of the $1/N_c$ expansion.
It is desirable to extend this analysis to higher orders of
$1/N_c$ expansion.
\par
On the other hand, there is another class of four-fermion
interaction, the Thirring model
\cite{Thirring58,Klaiber68}. This model is also important
as the low energy effective theory of QCD
\cite{ER86,AR95} as well as the field theoretical model
for studying the nonperturbative aspects of quantum field
theory, see references in 
\cite{IKSY95,Kondo95Th1,Kondo95Th2}.
The Thirring model is perturbatively
non-renormalizable in 3+1 dimensions. However, the 2+1
dimensional Thirring model is also renormalizable in
$1/N_f$ expansion, as well as the 2+1 dimensional NJL (or
Gross-Neveu) model.
In light of the above argument, it is interesting to pursue
the gauge-invariant generalization of the Thirring model,
which we call the {\it gauged Thirring model}.
Quite recently, the Thirring model was itself reformulated
as a gauge theory
\cite{IKSY95,Kondo95Th1,Kondo95Th2,IKN95}.   There it has
been realized that the gauge structure is actually
involved in the original Thirring model and the kinetic
term for the gauge field which is absent in the original
Lagrangian is generated through radiative corrections only
when the fermion is massive.  
In this paper we propose a gauged Thirring model. It
should be noted that the original Thirring model is
reproduced as the {\it strong gauge coupling limit} of the
gauged Thirring model, in sharp contrast with the gauged
NJL model in which the gauge interaction is introduced by
hand and the NJL model is obtained as the weak gauge
coupling limit. As the gauge-invariant generalization of
the four-fermion interaction, the gauged
Thirring model seems to be more natural than the gauged NJL
model.  In this paper we study the implication coming from
the gauge-invariant extension of the Thirring model,
especially paying attention to the effect of the kinetic
term for the gauge field.
\par
We also discuss the possible perturbatively
renormalizable extension of the gauged Thirring model.
It is easy to see that the gauged Thirring model is
equivalent to the gauged non-linear $\sigma$ model with
fermions, which implies that the gauged Thirring model is
perturbatively non-renormalizable.
We can extend the (perturbatively non-renormalizable)
gauged Thirring model into the perturbatively
renormalizable model, the generalized gauged Thirring
model.
However, it is not yet clear under what conditions the
gauged Thirring models give well-defined continuum limits
which are equivalent to specific nontrivial generalized
gauged Thirring models.
Such a viewpoint is quite necessary to consider the
extension of the work \cite{HKKN94} to the gauged Thirring
model and give an answer for the question raised in ref.
\cite{Hands95}.
\par
We have found that the recent proposal of
gauge-invariant extension of four-fermion model by Jersak
and his collaborators \cite{FJ95,FFJL95} is  essentially
the same as the gauged Thirring model which we have
proposed.  Their model defined on a lattice which they
call the $\chi U \phi$ model does not exactly coincides
with ours, since they use the compact
formulation for the gauge field and the staggered fermion
on the lattice.   However, the naive continuum limit of
$\chi U \phi$ model reduces to the gauged Thirring model,
if the staggered fermion is replaced with the Dirac
fermion.  Therefore, their extensive studies of Monte Carlo
simulation may shed more light on the non-perturbative
features of the gauged Thirring model from somewhat
different directions, although the direct comparison of
our result with theirs may not be possible.

\par
This paper is organized as follows.
The paper is divided into three parts. 
In part I, we give the
formulation of the gauged Thirring model.
In part II and III, we analyze the Abelian-gauged
Thirring model by making use of the effective potential
in part II and the Schwinger-Dyson (SD) equation in part
III. 
\par
In part I, section 2 and 3, the gauged Thirring model is
formulated.
In section 2, we define the gauged Thirring
model so that the Thirring model which is recently
reformulated as a gauge theory is recovered as the strong
gauge-coupling limit $\beta \rightarrow 0$.  The original
Thirring model is regarded as a gauge-fixed version of the
extended gauge theory, which we call the master gauge
theory. Moreover, we give the covariantly gauge-fixed BRST
formulation of the gauged Thirring model. To solve the
Schwinger-Dyson (SD) equation in part III, we need to take
the  nonlocal covariant gauge fixing \cite{GSC90,KM92}. 
It is shown that the above formulation can be extended to
the case of the nonlocal gauge fixing.
%\par
In section3, we restrict ourselves to the Abelian-gauged
Thirring model and discuss the limiting cases.  Based on
the considerations so far, we conjecture the possible phase
diagram of the gauged Thirring model.  In the Abelian
gauged case, we can integrate out the gauge boson field
exactly and can obtain the effective fermionic theory of
the gauged Thirring model, which reduces to the original
Thirring model in the strong gauge-coupling limit. 
This is nothing but the fermionization of the gauged
Thirring model. However, the effective fermionic theory
obtained in such a way is nonlocal and has no gauge
invariance. 
\par
In part II, from section 4 to 6,  we obtain the
effective potential for the chiral order parameter which
is used to obtain the phase diagram of the gauged Thirring
model.
%\par
Section 4 is a preparatory section to obtain the effective
potential.  In order to obtain the {\it gauge-independent}
effective potential, we adopt the inversion method
invented by Fukuda \cite{Fukuda} and developed with his
collaborators \cite{UKF90,Yokojima95,FukudaSuppl}. 
%\par
In section 5, we  explicitly obtain the effective
potential for the chiral order parameter according to the
inversion method.  It is used to show the existence of the
chiral phase transition which is the second order. 
%\par
In section 6, from the effective potential in the
previous section, we obtain the critical line of the
chiral phase transition in the phase diagram and determine
the critical exponent for the chiral order parameter.
\par
In part III, from section 7 to 12, we discuss the phase
structure in more detail and examine the continuum
limit based on the solution of the SD equation for the
gauged Thirring model. In section 7, we set up the SD
equation for the gauged Thirring model.  We argue the
consistency of the ladder approximation from the viewpoint
of gauge invariance.  This is necessary and also possible,
since the gauged Thirring model has now been formulated as
a true gauge theory. We show that the bare vertex
approximation is consistent with no wavefunction
renormalization for the fermion.  Such a  situation is
realized only when we take the nonlocal $R_\xi$ gauge.
%\par
In section 8, we obtain the Cornwall-Jackiw-Tomboulis
(CJT) effective action whose variation with respect to
a full propagator yields the SD equation for the full
propagator (in the nonlocal gauge).  We show that the
nontrivial (chiral-symmetry breaking) solution for the SD
equation gives lower effective potential than the trivial
(chiral symmetric) solution.  This implies that, in the
chiral limit, the vacuum favors the chiral symmetry
breaking rather than the chiral symmetric case.
\par
In section 9, we discuss the solution of the SD equation
in the nonlocal gauge.  We obtain the explicit solution in
two regions in the phase diagram which are neighborhood of
the two critical points of the pure Thirring$_4$ model and
the pure QED$_4$.
%\par
In section 10, we write the equi-correlation-length line
for the fermion based on the solution of SD equation.  The
critical line is obtained in the limit of infinite
correlation-length.  We discuss also the possible
existence of the solution in the negative four-fermion
coupling region.
%\par
In section 11, we study the RG flows (i.e. lines of
constant physics) for the constant renormalized coupling
constant and for the constant mass ratio between the
fermion mass and the gauge boson mass.
We examine the continuum limit of the gauged Thirring
model with cutoff and find the nontriviality of the
continuum gauged Thirring theory.
%\par
In section 12, the anomalous dimension for the
composite operator
$\bar \psi \psi$ is calculated in the gauged Thirring
model.  It is shown that the anomalous dimension of
the continuum gauged Thirring theory takes the large value
(greater than 1 and smaller than 2), if we require
the nontriviality of the continuum limit, namely, the
continuum limit is taken along the RG trajectories
obtained in the previous section. The final section is
devoted to conclusion and discussion.
\par
In Appendix A, we include the short note for the inversion
method which is sufficient to give the inversion formula.

\newpage
\section{Thirring model as a gauge theory}
\setcounter{equation}{0}
\par
We consider the general Thirring model whose
Lagrangian is given by
\begin{eqnarray}
 {\cal L}_{Th} 
 &=& \bar \psi^j i \gamma^\mu \partial_\mu \psi^j 
 - m_0 \bar \psi^j \psi^j
- {G_T \over 2N}
(\bar \psi^j  \gamma_\mu T^a \psi^j)
(\bar \psi^k  \gamma^\mu T^a \psi^k),
\label{NATh}
\end{eqnarray}
where $\psi^j$ is a 4-component Dirac fermion with an
flavor index $j$ which runs from 1 to $N_f$ (summed over
from 1 to $N_f$ if repeated) and
$\gamma^\mu (\mu=0,1,2,...,D-1)$ are $4 \times 4$ gamma
matrices satisfying the Clifford algebra 
$\{ \gamma_\mu, \gamma_\nu \} = 2 g_{\mu\nu} {\bf 1}$,
and $T^a (a=1,..., {\rm dim} G)$ makes up a basis of the
Lie group (color group) $G$, e.g. ${\rm dim} G=N_c^2-1$ for
$G=SU(N_c)$.
\par
By introducing the auxiliary vector field 
$A_\mu := g A_\mu^a T^a$, 
the theory is equivalently rewritten as
\begin{eqnarray}
 {\cal L}_{Th'}
 &=& \bar \psi^j i \gamma^\mu  D_\mu[A]  \psi^j 
 - m_0 \bar \psi^j \psi^j
 + {M^2 \over 2} (A_\mu^a)^2 ,
\label{NATh'}
\end{eqnarray}
where we have defined the covariant derivative:
\begin{eqnarray}
D_\mu[A] 
:=  \partial_\mu - i A_\mu
 = \partial_\mu - i g A_\mu^a T^a, 
\end{eqnarray}
and introduced another parameterization \cite{Kondo95Th2}:
\begin{eqnarray}
 G_T := {g^2 \over M^2}. 
\end{eqnarray}
\par
The theory given by the Lagrangian (\ref{NATh'}) is
identical with the massive Yang-Mills theory where the
Yang-Mills field couples minimally to fermions, although it
lacks the kinetic term for the Yang-Mills field.
Note that the kinetic term for the fermion and the fermion
mass term are gauge invariant, but only the mass term for
the vector field breaks the gauge invariance explicitly.
Therefore, the vector field $A_\mu$ is not still a gauge
field at this stage.

\subsection{recovery of gauge invariance}
As is well known from the study of massive Yang-Mills theory
\cite{Yokoyama81},
the theory can be rewritten into the gauge-invariant
form by introducing scalar modes 
$\theta^a (a=1,..., {\rm dim} G)$. Actually, introducing the
vector field
$K_\mu(\theta)$:
\begin{eqnarray}
K_\mu(\theta) = {i \over g} U_\theta \partial_\mu
U^\dagger_\theta  := K_\mu^a(\theta) T^a, 
\quad
 K_\mu^a(\theta) = {2i \over g} \tr[T^a U_\theta
\partial_\mu U^\dagger_\theta],
\label{defK}
\end{eqnarray}
made from an element of the unitary group $U(N_c)$:
\begin{eqnarray}
 U_\theta(x) = \exp [ig \theta^a(x) T^a] 
 := U^0(x) + i U^a(x) T^a,
\end{eqnarray}
we can write the gauge-invariant generalization of
Eq.~(\ref{NATh'}):
\begin{eqnarray}
 {\cal L}_{Th''} 
% &=& 
 = \bar \psi^j i \gamma^\mu D_\mu[A] \psi^j 
 - m_0 \bar \psi^j \psi^j
 +  M^2  \tr(A_\mu - K_\mu(\theta))^2 
%\nonumber\\&&
% - {\beta \over 2} \tr(F_{\mu\nu}F^{\mu\nu}),
\label{NATh''}
\end{eqnarray}
where we have used the following normalization:
\begin{eqnarray}
 \tr(T^a T^b) = {1 \over 2} \delta_{ab} .
\end{eqnarray}
\par
The scalar field $\theta := g \theta^a T^a$ introduced
above is nothing but the non-Abelian generalization of the
St\"uckelberg field 
\footnote{This field is also identified with the
Batalin-Fradkin (BF) field \cite{BF86} in the general
canonical Hamiltonian formalism of
Batalin-Fradkin-Vilkovisky \cite{BFV} (BFV) for the
constrained system, as pointed out in the previous paper
\cite{Kondo95Th1}, see the references \cite{FIK90} for
explicit derivation of Eq.~(\ref{NAgTh}) from this point
of view. }  
in the Abelian massive gauge theory
\cite{Thirring58}.
Indeed, the Lagrangian Eq.~(\ref{NATh''}) is invariant
under the (finite) local gauge transformation:
\begin{eqnarray}
% \psi^j(x) &\longmapsto& 
\psi_j '(x) &=& U_\omega(x) \psi_j(x),
 \nonumber\\
% A_\mu(x) &\longmapsto& 
 A_\mu '(x) &=& U_\omega(x) A_\mu(x)U_\omega^\dagger(x)
 + {i \over g} U_\omega(x) \partial_\mu U_\omega^\dagger(x)
 \nonumber\\
 &=& U_\omega(x) \left [A_\mu(x) - {i \over g}
U_\omega^\dagger(x)
\partial_\mu U_\omega(x) \right] U_\omega^\dagger(x),
 \nonumber\\
% K_\mu(x) &\longmapsto& 
 K_\mu '(x) &=& 
 {i \over g} U'(x) \partial_\mu U'(x)^\dagger
 \quad (U'(x) := U_\omega(x)U_\theta(x))
 \nonumber\\
 &=& U_\omega(x) K_\mu(x)U_\omega^\dagger(x) 
 + {i \over g} U_\omega(x) \partial_\mu U_\omega^\dagger(x),
 \nonumber\\
 &=& U_\omega(x) \left[ K_\mu(x) - {i \over g}
U_\omega^\dagger(x)
\partial_\mu U_\omega(x) \right] U_\omega^\dagger(x) .
\end{eqnarray}
\par
>From the reason explained in the introduction, we further
introduce the kinetic term for the {\it gauge} field
$A_\mu$ which is itself gauge-invariant:
\begin{eqnarray}
 - {\beta \over 2} \tr(F_{\mu\nu}F^{\mu\nu}),
\end{eqnarray}
where the field strength is defined by
$
F_{\mu\nu} := g  F_{\mu\nu}^a T^a
= \partial_\mu A_\nu - \partial_\nu A_\mu
- ig[A_\mu, A_\nu]
$ 
and 
$
[T^a, T^b] = if_{abc} T^c
$
with the structure constant $f_{abc}$ of $G$.
Thus we arrive at a gauge-invariant generalization of the
Thirring model:
\footnote{
Here we have rescaled the gauge field
$A_\mu^a$ and the scalar field $\theta^a$ by $g^{-1}$.
Hence, in what follows, $M^2$ should be identified with
$G_T^{-1}$: $M^2=G_T^{-1}$.
}
\begin{eqnarray}
 {\cal L}_{gTh} 
 &=& \bar \psi^j i \gamma^\mu D_\mu[A] \psi^j 
 - m_0 \bar \psi^j \psi^j
 +  G_T^{-1}  \tr(A_\mu - K_\mu(\theta))^2 
\nonumber\\&&
 - {\beta \over 2} \tr(F_{\mu\nu}F^{\mu\nu}),
 \quad
 \beta := {1 \over g^2}.
\label{NAgTh}
\end{eqnarray}
The $\beta=0$ case is the Thirring model. 
We call the model given by the Lagrangian
Eq.~(\ref{NAgTh}) with non-zero $\beta$ the {\it gauged
Thirring model}.
\par
The original Thirring model is identified with a
gauge-fixed version of the gauge theory with the
Lagrangian, Eq.~(\ref{NAgTh}), which we call the
{\it master gauge theory}.  
\footnote{
The master gauge theory is not unique.
Actually we can write another type of the master gauge
theory for the Thirring model \cite{IKN95}, which is
easily extended to the gauged Thirring model. 
This type of the master gauge theory plays the important
role of the bosonization of the Thirring model.
}  
Indeed, the Lagrangian Eq.~(\ref{NAgTh}) reduces to the
Lagrangian Eq.~(\ref{NATh'}), if we take the unitary gauge:
$\theta(x) \equiv 0$ (and $\beta=0$).
However, actual calculations such as loop calculations are
generally impossible in the unitary gauge.  
For such purposes the covariant gauge is most convenient,
although both gauge-fixing should give the same results on
the gauge-invariant quantities, e.g., the chiral condensate
$\langle \bar \Psi \Psi \rangle$.
Various advantages of maintaining such a gauge symmetry
were emphasized in the previous works \cite{IKSY95}.
\par
The theory with the Lagrangian ${\cal L}_{Th''}$ can be
cast into the form which is similar to the
Higgs-Kibble model (or gauged non-linear sigma model) in
the presence of fermions: 
\begin{eqnarray}
 {\cal L}_{HK} 
 &=& \bar \psi^j i \gamma^\mu  D_\mu [A]  \psi^j 
 - m_0 \bar \psi^j \psi^j
 \nonumber\\&&
 +  G^{-1} \tr[(D_\mu[A] \Phi)^\dagger (D^\mu[A] \Phi)]
  - {\beta \over 2} \tr(F_{\mu\nu}F^{\mu\nu}) .
\end{eqnarray}
with a local constraint: 
\begin{eqnarray}
\Phi(x) \Phi^\dagger(x) \equiv 1,
\label{constraint}
\end{eqnarray}
for the scalar field $\Phi(x) := U(x)$.
It should be noted that the theory with this Lagrangian is
perturbatively non-renormalizable, as well as the original
Thirring model.
\par
In the chiral limit $m_0 \rightarrow 0$, the gauged
Thirring model as well as the original Thirring model has
the chiral-symmetry, namely, the Lagrangian is invariant
under the chiral transformation:
\begin{eqnarray}
 \psi^j(x) &\longmapsto& 
\psi_j '(x) := e^{i \gamma_5 \theta} \psi_j(x),
 \nonumber\\
 \bar \psi^j(x) &\longmapsto& 
\bar \psi_j '(x) := 
\bar \psi_j(x) e^{i \gamma_5 \theta}  .
\end{eqnarray}

\subsection{BRST quantization}
\par
In the following of this paper, we adopt the covariant
gauge fixing for the gauged Thirring model.
In what follows, therefore, we give the covariantly
gauge-fixed BRST formulation of the gauged Thirring model.
It is well known that the total Lagrangian 
\begin{eqnarray}
 {\cal L}_{Th'''} = {\cal L}_{Th''} + {\cal L}_{GF} 
+ {\cal L}_{FP},
\end{eqnarray}
which is invariant under the BRST transformation $\delta_B$
is obtained by adding the gauge-fixing term and the
Faddeev-Popov (FP) ghost term
${\cal L}_{GF+FP}$ to the Lagrangian ${\cal L}_{Th''}$:
\begin{eqnarray}
 {\cal L}_{GF+FP}[A, \theta, C, \bar C,B]
 &=& - i 
 \delta_B(\bar C^a (F^a[A,\theta]-{\xi \over 2}B^a)),
\end{eqnarray}
where $\delta_B$ denotes the BRST transformation.
The nilpotent BRST transformation $\delta_B^2 * = 0$ is
given by
\begin{eqnarray}
 \delta_B A_\mu^a(x) &=&  D_\mu^{ab}(A) C^b(x),
\nonumber\\ 
 \delta_B B^a(x) &=& 0,
\nonumber\\ 
 \delta_B C^a(x) &=& -  {g \over 2} f^{abc}C^b(x)
C^b(x),
\nonumber\\ 
 \delta_B \bar C^a(x) &=& - i  B^a(x),
\nonumber\\ 
 \delta_B \psi^j(x) &=&  i  g  C^a(x) T^a \psi^j(x),
\nonumber\\ 
 \delta_B K_\mu^a(x) &=&  D_\mu^{ab}(K) C^b(x),
 \label{BRST}
\end{eqnarray}
where
\begin{eqnarray}
 D_\mu^{ab}[A]  := \delta_{ab}
 \partial_\mu - i g f^{abc} A^c_\mu .
\end{eqnarray}
Note that the BRST transformation of $K_\mu$ can be
rewritten into that of $\theta$:
\begin{eqnarray}
 \delta_B \theta^a(x) &=&  K^{-1}(\theta)^{ab} C^b,
\end{eqnarray}
where $K^{-1}(\theta)^{ab}$ is the inverse of
$K(\theta)^{ab}$ which is defined through
\begin{eqnarray}
 K_\mu^a(\theta) = K^{ab}(\theta) \partial_\mu \theta^b .
\end{eqnarray}
Note that the BRST invariance of the total Lagrangian is
guaranteed, since
${\cal L}_{GF+FP}$ is invariant under the BRST
transformation due to nilpotency of BRST
transformation Eq.~(\ref{BRST})  and the original
Lagrangian ${\cal L}_{Th''}$ is also invariant. 
\par
For the BRST transformation Eq.~(\ref{BRST}), we have
\begin{eqnarray}
 {\cal L}_{GF+FP}
 &=& B^a F^a[A,\theta]+{\xi \over 2} (B^a)^2
 - i \bar C^a \delta_B F^a[A,\theta] .
\end{eqnarray}
After eliminating the $B$ field, we obtain
\begin{eqnarray}
 {\cal L}_{GF}
 &=& - {1 \over 2\xi} (F^a[A,\theta])^2,
 \label{GF}
 \\
{\cal L}_{FP} &=& - i \bar C^a \delta_B F^a[A,\theta] 
 \nonumber\\
&=& - i \bar C^a \left( {\delta F^a[A,\theta]
\over \delta A_\mu^c} D_\mu^{ab}[A] C^b(x) 
+ {\delta F^a[A,\theta] \over \delta \theta^c}
K^{-1}(\theta)^{ab} C^b  \right).
\end{eqnarray}

\subsubsection{a covariant gauge}
\par
The conventional covariant gauge-fixing is obtained for the
following
$F^a$: 
\begin{eqnarray}
 F^a[A,\theta] = \partial^\mu A_\mu^a .
\end{eqnarray} 
This covariant gauge leads to the following ghost term:
\begin{eqnarray}
 {\cal L}_{FP}
 =  - i \bar C^a  D_\mu^{ab}[A] C^b(x) .
\end{eqnarray}
In this covariant gauge, we notice that there is a crossing
term 
$A_\mu K_\mu(\theta)$ in the Lagrangian ${\cal L}_{Th''}$.
This term sometimes makes the analysis somewhat
cumbersome.

\subsubsection{$R_\xi$ gauge}
\par
We consider another choice for the gauge-fixing function.
In the following we choose the generalized
$R_\xi$ gauge:
\begin{eqnarray}
 F^a[A,\theta] = \partial^\mu A_\mu^a + \xi f^a(\theta),
\end{eqnarray} 
such that the crossing term $A_\mu K_\mu(\theta)$ is
canceled with the relevant term in the gauge-fixing
term:
\begin{eqnarray}
 {\cal L}_{GF}
 = - {1 \over 2\xi} (F^a[A,\theta])^2 
 = - {1 \over 2\xi} (\partial^\mu A_\mu^a)^2 
 - f^a(\theta) \partial^\mu A_\mu^a 
 - {\xi \over 2} (f^a(\theta))^2,
\end{eqnarray}
in the total Lagrangian 
${\cal L}_{Th'''}$.
For such a cancellation to occur, we must choose
$f^a$ such that $f^a$ satisfies the following differential
equation:
\begin{eqnarray}
  \partial^\mu f^a(\theta) = M^2 K_\mu^a(\theta)
  = M^2 K^{ab}(\theta) \partial_\mu \theta^b .
  \label{gc}
\end{eqnarray}
Instead, the $R_\xi$ gauge generates a new type of
interaction between the ghost field and the scalar mode in
the FP ghost term:
\begin{eqnarray}
 {\cal L}_{FP}[A, \bar C, C, \theta]  
 = - i \bar C^a  \partial^\mu D_\mu^{ab}(A) C^b  
 - i \bar C^a  \xi {\delta f^a(\theta)
\over \delta \theta^c} K^{-1}(\theta)^{cb} C^b .
\end{eqnarray}
\par
In the Abelian case, $K^{ab}=\delta_{ab}$, we have
\begin{eqnarray}
 K_\mu(\theta) = \partial_\mu \theta .
\end{eqnarray}
Then the equation Eq.~(\ref{gc}) is easily integrated to
give the solution:
\begin{eqnarray}
  f(\theta) = M^2 \theta .
\end{eqnarray}
This recovers the usual $R_\xi$ gauge 
\begin{eqnarray}
F[A,\theta]=\partial^\mu A_\mu + \xi M \theta .
\end{eqnarray}
In the Abelian case, the total Lagrangian is decomposed
as
\cite{IKSY95}:
\begin{eqnarray}
 {\cal L}_{Th'''} 
 &=& {\cal L}_{\psi,A} 
+ {\cal L}_{\theta}
+ {\cal L}_{FP} ,
\nonumber\\
{\cal L}_{\psi,A}
&=& \bar \psi^j i \gamma^\mu D_\mu[A]\psi^j 
 - m_0 \bar \psi^j \psi^j
 + {M^2 \over 2}(A_\mu^a)^2
 + {1 \over 2\xi}(\partial^\mu A_\mu)^2
  - {\beta \over 4}  F_{\mu\nu}F^{\mu\nu}  .
\nonumber\\
{\cal L}_{\theta}
&=&  {1 \over 2G_T}(\partial_\mu \theta)^2
 - {\xi \over 2} (f(\theta))^2,
\nonumber\\
 {\cal L}_{FP} &=& {\cal L}_{FP}[\bar C, C] 
= - i \bar C \left( \partial_\mu \partial^\mu
 + {\xi \over G_T} \right) C .
 \label{abeliangTh}
\end{eqnarray}
In the Abelian case, therefore, the scalar mode
can be decoupled for an arbitrary $\xi$ by choosing the
$R_\xi$ gauge,
since the FP ghost term ${\cal L}_{FP}$ does not involve
the $\theta$ field.
\footnote{
Note that ${\cal L}_{\psi,A}$ is aparantly  equivalent to
the starting Lagrangian of Gomes et al. \cite{GMRS91}.
However,  ${\cal L}_{\psi,A}$ is a natural consequence of
the full gauge-invariant formulation of the (gauged)
Thirring model and its meaning is completely different from
theirs. }
\par
In the non-Abelian case, also, the equation Eq.~(\ref{gc})
may have the solution, since $K_\mu$ defined by
Eq.~(\ref{defK}) satisfies the integrability condition
\cite{Yokoyama81}:
\begin{eqnarray}
 K_{\mu\nu}^a \equiv
 \partial_\mu K_\nu^a - \partial_\mu K_\nu^a 
 + g f^{abc} K_\mu^b K_\nu^c = 0,
\end{eqnarray} which is nothing but the pure-gauge
condition (or a null-curvature equation). 
Assuming that $f^a(\theta)$ is expressed in terms of
$\theta$ by solving Eq.~(\ref{gc}), we can write the total
Lagrangian:
\begin{eqnarray}
 {\cal L}_{Th'''} 
 &=& \bar \psi^j i \gamma^\mu D_\mu[A] \psi^j 
 - m_0 \bar \psi^j \psi^j
\nonumber\\&&
 - {\beta \over 2} \tr(F_{\mu\nu}F^{\mu\nu})
 + {M^2 \over 2}(A_\mu^a)^2
+ \tilde B^a  (\partial^\mu A_\mu^a) 
+ {\xi \over 2} (\tilde B^a)^2
\nonumber\\&&
 + {M^2 \over 2}(K_\mu^a(\theta))^2
 - {\xi \over 2} (f^a(\theta))^2
 + {\cal L}_{FP}[A, \bar C, C, \theta] ,
\end{eqnarray}
where we have introduced a new Lagrange multiplier field 
$\tilde B^a(x)$
\par
In the Non-Abelian case,  
the scalar mode $\theta^a$ is  not decoupled, unless we
impose further condition.  
Then, in the non-Abelian case, the self-interacting scalar
field is decoupled only in the very special gauge.
More details on the non-Abelian case will be given in
subsequent papers \cite{Kondo96}.

\subsection{nonlocal gauge fixing}

In order to solve the SD equation in the ladder
approximation in section 9, we need to take the nonlocal
gauge. Here the nonlocal gauge is used in the sense that the
gauge-fixing parameter $\xi$ gets momentum-dependent, i.e.
$\xi$ becomes a function of the momentum:
$\xi=\xi(k^2)$.
In the configuration space, the gauge fixing term in the
nonlocal gauge is given by
\begin{equation}
  {\cal L}_{GF}
  = - {1 \over 2} F^a[A(x), \theta(x)] \int d^D y
  {1 \over \xi(x-y)}F^a[A(y), \theta(y)],
\label{nlgfterm}
\end{equation}
which has been derived for the covariant gauge
\cite{KM95} and the extension to the $R_\xi$ gauge is
straightforward, see e.g. \cite{IKSY95}. 
Here it should be noted that  $\xi^{-1}(k^2)$ is the
Fourier transform of $\xi^{-1}(x)$, 
\begin{equation}
  \xi^{-1}(x) = \int {d^Dk \over (2\pi)^D}
e^{ikx} \xi^{-1}(k^2),
\quad
  \xi^{-1}(k^2) = \int  d^D x
e^{-ikx} \xi^{-1}(x),
\end{equation}
while $\xi(k^2)$ is not the Fourier
transform of $\xi(x)$, see ref.~\cite{KM95}.
If $\xi(k^2)$ does not have the
momentum-dependence, i.e.,
$\xi(k^2) \rightarrow \xi$, then 
$\xi^{-1}(x-y) \rightarrow \delta(x-y)\xi^{-1}$ and the
nonlocal gauge-fixing term reduces to the usual
gauge-fixing term, Eq.~(\ref{GF}).
In the formal way, the BRST formulation can be
easily extended into the case of nonlocal gauge-fixing by
replacing the product with the convolution when
the gauge-fixing function $\xi(x)$ appears, as shown
explicitly for the Abelian case in
\cite{IKSY95}.

\subsection{renormalizable extension}
I want to regard the gauged Thirring model with a
special limit of the extended theory which is  
perturbatively renormalizable.  The simplest way is to
add the potential term of the scalar field $\Phi$ into the
total Lagrangian:
\begin{eqnarray}
 {\cal L}_{pot} 
 &=& \mu^2 \tr(\Phi^\dagger \Phi)
 + \lambda (\tr(\Phi^\dagger \Phi))^2 ,
\end{eqnarray}
at the same time of removing the constraint
Eq.~(\ref{constraint}). The theory with the Lagrangian 
${\cal L}_{Th''}+{\cal L}_{pot}$ 
is equivalent to the Higgs-Kibble model with fermions.
Therefore, it is quite interesting to study the
relationship between the gauged Thirring model and the
renormalizable extension.
This will be done in a subsequent paper \cite{Kondo96}.

\newpage
\section{Abelian-gauged Thirring model}
\setcounter{equation}{0}

\subsection{reformulation as a gauge theory}
In this paper we study in detail the Abelian-gauged Thirring
model with the following Lagrangian:
\begin{eqnarray}
 {\cal L}  
 =  \bar \psi^j (i \gamma^\mu \partial_\mu - m_0) \psi^j
 - A_\mu \bar \psi^j \gamma^\mu \psi^j
 + {1 \over 2G_T} |(\partial_\mu + i A_\mu) \phi |^2
 - {\beta \over 4} F^{\mu\nu}F_{\mu\nu} ,
 \label{gTh}
\end{eqnarray}
where $j = 1, ..., N_f$ and $\phi$ is the complex scalar
field of unit length $|\phi(x)|=1$ parameterized as
\begin{eqnarray}
 \phi(x) = e^{i \theta(x)}.
\end{eqnarray}
By using the variable $\theta$, this model is equivalently
rewritten into the following one:
\begin{eqnarray}
 {\cal L}  
 =  \bar \psi^j (i \gamma^\mu \partial_\mu - m_0) \psi^j
 - A_\mu \bar \psi^j \gamma^\mu \psi^j
 + {1 \over 2G_T} (A_\mu - \partial_\mu \theta)^2
 - {\beta \over 4} F^{\mu\nu}F_{\mu\nu} .
 \label{gTh'}
\end{eqnarray}
This theory is invariant under the local $U(1)$
gauge transformation:
\begin{eqnarray}
 \psi(x) &\rightarrow& \psi(x) e^{i \omega(x)}.
 \nonumber\\
 A_\mu(x) &\rightarrow& A_\mu(x) + \partial_\mu \omega(x),
 \nonumber\\
 \phi(x) &\rightarrow& \phi(x) e^{i \omega(x)}
\quad (\theta(x) \rightarrow \theta(x) + \omega(x)),
 \label{gt}
\end{eqnarray}
Therefore we need a gauge-fixing term to quantize this
theory as discussed in the previous section.

\subsection{limiting cases}
\par
In the strong gauge-coupling limit $\beta \rightarrow 0$,
this model reduces to the gauge invariant-reformulation of
the (massive) Thirring model
where $\theta$ is nothing but the well-known
St\"uckelberg field. Indeed, after integrating out the
bosonic field
$A_\mu$ under the unitary gauge $\theta=0$, we arrive at
the Lagrangian of the Thirring model:
\begin{eqnarray}
 {\cal L}  
 =  \bar \psi^j (i \gamma^\mu \partial_\mu - m_0) \psi^j
 - {G_T \over 2} (\bar \psi^j \gamma^\mu \psi^j)^2 .
\end{eqnarray}
Therefore the Thirring model is identified with a
gauge-fixed version of the master gauge theory
(\ref{gTh}) or (\ref{gTh'}).  From this point of view, it
is possible
\cite{Kondo95Th1} to regard the field $\theta$ as the
Batalin-Fradkin field
\cite{BF86} in the general theory of the constrained
Hamiltonian system
\cite{BFV}. In the case of $\beta = 0$, the field $A_\mu$
is an auxiliary vector field which is introduced so as to
linearize the four-fermion interaction.  Hence it does not
have the corresponding kinetic term.   
In our model we have introduced the kinetic term, 
${\beta \over 4} F^{\mu\nu}F_{\mu\nu}$ from the beginning,
after the identification of the field $A_\mu$ with the
gauge field.  Therefore, at least for non-zero $\beta$, the
field $A_\mu$ is no longer auxiliary but is regarded as
true gauge field which obeys the gauge transformation,
Eq.~(\ref{gt}). 
\par
Even in the case of $\beta=0$, it has been
shown that the kinetic term for the field $A_\mu$ is
generated due to radiative corrections
\cite{Hands95}, only when the fermion mass is
dynamically generated in the chiral limit $m_0=0$.  
In the case of massive fermion, particularly very massive
fermion, the generation of the kinetic term can be seen also
through the procedure of bosonization
\cite{FS94,Kondo95Th2,IKN95}.
To see the effect coming from the dynamical gauge field
$A_\mu$, we have here introduced the
kinetic term from the beginning.
This corresponds to enlarging the bare parameter space in
accord with the concept of Wilsonian RG.
Our model is regarded as a gauge-invariant generalization of
the Thirring model where all the fields have the
corresponding kinetic term. We call this model the
(Abelian-) gauged Thirring model. Such a proposal, i.e.,
gauge-invariant generalization of the four-fermion model
has been given quite recently on the lattice by Jersak et
al. \cite{FJ95}. The correspondence of this model to ours
will be discussed in the final section.
\par
In the limit 
$1/G_T \rightarrow 0 (G_T \rightarrow \pm \infty)$, 
the gauged Thirring model reduces to the quantum
electrodynamics (QED) with $N_f$ flavors of fermions:
\begin{eqnarray}
 {\cal L}_{QED}  
 =  \bar \psi^j (i \gamma^\mu \partial_\mu - m_0) \psi^j
 - A_\mu \bar \psi^j \gamma^\mu \psi^j
 - {\beta \over 4} F^{\mu\nu}F_{\mu\nu} 
 \label{qed}
\end{eqnarray}
\par
In the weak gauge-coupling limit $\beta \rightarrow \infty$,
on the other hand, the gauged Thirring model approaches to
the non-linear
$\sigma$ model:
\begin{eqnarray}
 {\cal L}_{NL\sigma}  =  {1 \over 2G_T}
 |\partial_\mu  \phi |^2   \quad (|\phi | = 1),
 \label{nls}
\end{eqnarray}
since in this limit the gauge field reduces to the pure
gauge, $A_\mu = \partial_\mu \vartheta$ and $\vartheta$ can
be absorbed into the scalar and the fermion field and
hence the fermionic part reduces to the free theory.

\subsection{order parameter and phase transition}
\par
Here we mention the order parameter which is used to probe
the phase transition of the gauged Thirring model.
In the massless fermion, $m_0=0$, the chiral condensation
$\langle \bar \psi \psi \rangle$
can be used to distinguish the
spontaneous-chiral-symmetry-breaking phase from the
chiral-symmetric phase.  For the chiral-symmetry-breaking
phase transition, we know that there exists one
(non-zero and finite) critical coupling $\kappa_c$ in the
Thirring model as well as the coupling constant
$\beta_c$ in QED$_4$. Therefore
it will be natural to imagine that there will exist a
critical line which connects two critical points and the
chiral symmetry is broken spontaneously below
this critical line in the phase diagram
$(\beta,G_T)$.
The expected phase diagram is schematically shown in
Fig.~1. In this paper we will confirm this picture and
obtain the explicit form of this critical line based on the
effective potential for the order parameter, 
$\langle \bar \psi \psi \rangle$ as well as the solution
of the SD equation for the fermion propagator. At the same
time, we determine the order of the chiral phase transition
together with the critical exponent.

\par
\subsection{gauge fixing and fermionic action}
In the following we discuss the behavior of the
Lorentz-covariant and gauge-invariant order parameter,
especially the chiral condensate.  Since it is gauge
invariant, it should not depend on which type of covariant
gauge-fixing we may adopt in the quantization. In this paper
we choose the most convenient gauge in performing
the calculation.
In the previous section, we have shown that, 
if we choose the $R_\xi$ gauge, the St\"uckelberg field
$\theta$ in the gauged Thirring model is completely
decoupled irrespective of the value of the gauge-fixing
parameter $\xi$. 
Then the total Lagrangian ${\cal L}_{tot}$ is decomposed
into ${\cal L}_{\psi,A}$ and ${\cal L}_{\theta}$, see
Eq.~(\ref{abeliangTh}).  The Lagrangian ${\cal L}_{\psi,A}$
is rewritten as
\begin{eqnarray}
 {\cal L}_{\psi,A} &=&
\bar \psi^j(x) (i \gamma^\mu \partial_\mu - m_0) \psi^j(x)
 + A_\mu (x) \bar \psi^j(x)  \gamma^\mu \psi^j(x)
\nonumber\\&& + {1 \over 2}  
 A^\mu(x) D^{(0)}_{\mu\nu}{}^{-1}(x) A_\nu(x),
\end{eqnarray}
where
\begin{eqnarray}
 D^{(0)}_{\mu\nu}{}^{-1}(x)  
 = (-\beta \partial^\mu \partial_\mu + M^2) g_{\mu\nu}
+  (\beta + \xi^{-1}) \partial_\mu \partial_\nu . 
\end{eqnarray}
This is the inverse of the bare gauge boson propagator in
the $R_\xi$ gauge.  In momentum space it reads
\begin{eqnarray}
 D^{(0)}_{\mu\nu}{}^{-1}(k)  
 &=& (\beta k^2 + M^2) g_{\mu\nu}
-  (\beta + \xi^{-1}) k_\mu k_\nu 
\nonumber\\
 &=&  \beta k^2 \left( g_{\mu\nu} - {k_\mu k_\nu
\over k^2} \right) 
 + M^2 g_{\mu\nu}
-   \xi^{-1} k_\mu k_\nu . 
\end{eqnarray}
In the Abelian gauge theory, we can completely integrate
out the gauge field $A_\mu$.  Then we get the effective
fermionic action \footnote{The summation over the spacetime point is
understood if it is repeated.}
of the Abelian-gauged Thirring model:
\begin{eqnarray}
 S_{\psi}
=  \bar \psi^j(x) (i \gamma^\mu \partial_\mu - m_0)
\psi^j(x)
 + {1 \over 2}\bar \psi^j(x)  \gamma^\mu \psi^j(x)  
D^{(0)}_{\mu\nu}(x-y) 
\bar \psi^k(y)  \gamma^\nu \psi^k(y),
\label{effgTh}
\end{eqnarray}
where the gauge boson propagator is obtained as
\begin{eqnarray}
D^{(0)}_{\mu\nu}(x) 
&=& \int {d^D k \over (2\pi)^D} e^{ikx}
D^{(0)}_{\mu\nu}(k),   
\nonumber\\
D^{(0)}_{\mu\nu}(k) &=&  {1 \over \beta k^2 + M^2} 
\left(g_{\mu\nu}  - {k_\mu k_\nu \over k^2} \right)
+ {\xi \over k^2 + \xi M^2} {k_\mu k_\nu \over k^2}.
\label{gbpropa}
\end{eqnarray}
In the strong gauge-coupling limit $\beta=0$, we have
\begin{eqnarray}
D^{(0)}_{\mu\nu}(k) 
=  {1 \over M^2} \left( g_{\mu\nu} - 
{k_\mu k_\nu \over k^2 + \xi M^2} \right).
\end{eqnarray}
Then it turns out that the gauged Thirring model in the
$R_\xi$ gauge  reproduces the original Thirring model in
the limit $\xi \rightarrow \infty$   (namely, the unitary
gauge $\theta \rightarrow 0$), 
\footnote{If we require the current conservation:
$\partial_\mu (\bar \psi(x)  \gamma^\mu \psi(x)) = 0$,
the fermionic action does not depend on the gauge-fixing
parameter.}
since
\begin{eqnarray}
 D^{(0)}_{\mu\nu}(x-y) \rightarrow M^{-2}
g_{\mu\nu} \delta^D(x-y) = G_T g_{\mu\nu} \delta^D(x-y) .
\end{eqnarray}
On the other hand, in the weak gauge coupling limit $\beta
\rightarrow +\infty$, $S_0$ reduces to the action of the
free fermion.
Note that the effective fermionic theory given by
Eq.(\ref{effgTh}) is nonlocal and has no gauge invariance,
while the chiral symmetry is preserved in the chiral limit.

\subsection{renormalizable extension}
\par
It is interesting to consider more general
model where the scalar field is not fixed to be of a fixed
length:
\begin{eqnarray}
 {\cal L}  
 &=&  \bar \psi^a (i \gamma^\mu \partial_\mu - m_0) \psi^a
 - A_\mu \bar \psi \gamma^\mu \psi
 - {\beta \over 4} F^{\mu\nu}F_{\mu\nu} 
 \nonumber\\
&& + {1 \over 2G_T} |(\partial_\mu + i A_\mu) \phi |^2
 - m_\phi^2 |\phi |^2 - {\lambda \over 6} (|\phi |^2)^2,
 \label{egTh}
\end{eqnarray}
\par
Separating the scalar field as follows,
\begin{eqnarray}
 \phi(x) = {1 \over \sqrt{2}}\rho(x) e^{i \theta(x)},
\end{eqnarray}
we have
\begin{eqnarray}
 |(\partial_\mu - i A_\mu) \phi |^2
 &=&  {1 \over 2} (\partial_\mu \rho)^2 
 + {1 \over 2} \rho^2 (\partial_\mu \theta)^2
 + {1 \over 2} \rho^2 A_\mu A^\mu -  \rho^2 A_\mu
\partial^\mu \theta .
\end{eqnarray}
In the unitary gauge $\theta=0$, this term reads
\begin{eqnarray}
 |(\partial_\mu - i A_\mu) \phi |^2
 &=&  {1 \over 2} (\partial_\mu \rho)^2  
 + {1 \over 2} \rho^2 A_\mu A^\mu  ,
\end{eqnarray}
If we fix the length of the scalar field: $\rho(x)=\rho_0$,
then we have
\begin{eqnarray}
 |(\partial_\mu - i A_\mu) \phi |^2
 &=& {1 \over 2} \rho_0^2 (\partial_\mu \theta)^2
 + {1 \over 2} \rho_0^2 A_\mu A^\mu 
 -  \rho_0^2 A_\mu \partial^\mu \theta,
\end{eqnarray}
This is a renormalizable extension of the gauged Thirring
model.  Therefore, it is interesting to perform the
renormalization group (RG) analysis of the model and study
the relationship with this model and the gauged Thirring
model. This will enable us to extend the phase structure of
the gauged Thirring model and consider the universality
between this model and the Thirring model based on the
gauge-invariant formulation. The gauged Thirring model is
obtained as the limit
$\lambda \rightarrow \infty$ from this model by adjusting
the bare mass parameter $m_\phi^2$ appropriately. Hence the
usual perturbation theory with respect to the coupling
constant $\lambda$ is insufficient to study the gauged
Thirring model and we need some non-perturbative
methods.
This issue will be discussed in subsequent papers.

\newpage
\section{Chiral order parameter}
\setcounter{equation}{0}

In the following three sections, we pay attention to the
chiral order parameter 
\begin{eqnarray}
 \langle \bar \psi \psi  \rangle  
:= {1 \over N_f} \sum_{j=1}^{N_f}
\langle \bar \psi^j(x) \psi^j(x)  \rangle   
\end{eqnarray}
and obtain finally the effective
potential in terms of the chiral order parameter
according to the inversion method
\cite{Fukuda,UKF90,Yokojima95,FukudaSuppl}, see Appendix. 
This order parameter is gauge-invariant.  Therefore this
quantity should not depend on which type of gauge fixing we
may choose.
 In the following, we choose the
$R_\xi$ gauge which is the simplest gauge to calculate the
chiral order parameter in the gauged Thirring model, since
in this gauge the scalar field is completely decoupled
irrespective of the gauge-fixing parameter.
\par
In order to evaluate the chiral order parameter and
discuss the spontaneous chiral symmetry breaking in the
chiral limit $m_0=0$, we introduce the following
local source term into the Lagrangian according to the
rule of the inversion method:
\begin{eqnarray}
 {\cal L}_{J}  =  J \bar \psi^j(x)  \psi^j(x) .
\end{eqnarray}
\par
It is known that the vacuum expectation
value of the fermion composite operator $\bar \psi(x)
\psi(x)$ is calculated according to the expression:
\begin{eqnarray}
 \langle \bar \psi \psi  \rangle  
= \int {d^Dk \over (2\pi)^D} {\tr(1)J \over
p^2+J^2} + {1 \over 2N_f} (D-1)\int {d^Dk \over (2\pi)^D}
\left[
 k^2 D_T(k)  {\partial \over \partial J} 
 \Pi(k) \right],
 \label{cop}
\end{eqnarray}
if we use the following form of the gauge boson propagator:
\begin{eqnarray}
 D_{\mu\nu}(k)  =  D_T(k) 
\left(g_{\mu\nu}  - {k_\mu k_\nu \over k^2} \right)
+ D_L(k) {k_\mu k_\nu \over k^2}.
\end{eqnarray}
This relation shows that the chiral condensation can be
calculated irrespective of the longitudinal part of the
gauge boson propagator.  
This implies that the chiral order parameter obtained
according to Eq.~(\ref{cop}) does not depend on the
explicit form of $D_L$ and hence the gauge-fixing
parameter $\xi$.
This is expected from the gauge invariance of the chiral
condensate.  However, in the usual calculation, a specific
gauge is chosen for calculating the gauge-invariant
quantity, see e.g. \cite{SW92}.  The origin of this
difference has been explained in ref.~\cite{Kondo95MCS}.
This form has been already used to analyze the gauged
NJL and the gauged Yukawa models \cite{Kondo93}, 
the Thirring model in $D$ ($4>D>2$) dimensions
\cite{Kondo95Th1} and the three-dimensional QED
\cite{Kondo95MCS}, in order to obtain the gauge-invariant
phase structure.
\par
The first part of the right-hand-side (RHS) of
Eq.~(\ref{cop}) is obtained from the following relation for
the fermion propagator $S$: 
\begin{eqnarray}
 {\partial \over \partial J}
 \ln \det [S_0^{-1}]
=  {\partial \over \partial J}
  \tr \ln [S_0^{-1}]
=    \tr [S_0]
=    \int {d^Dk \over (2\pi)^D} {\tr(1)J \over p^2+J^2},
\label{fpart}
\end{eqnarray}
which comes from the part:
\begin{eqnarray}
\ln \int [d\bar \psi] [d\psi] \exp \left[ \int d^D x
\bar \psi S_0^{-1}\psi \right]
= \ln \det S_0^{-1},
\label{fgenerating}
\end{eqnarray}
where $S_0$ is the bare fermion propagator in the presence
of source term:
\begin{eqnarray}
 S_0(p)  = (\gamma^\mu p_\mu + J)^{-1}. 
\end{eqnarray}
\par
The second part of the RHS of Eq.~(\ref{cop})
has an alternative expression: 
\begin{eqnarray}
\int {d^D k \over (2\pi)^D} D_{\mu\nu}(k) 
{\partial \over \partial J} \Pi^{\mu\nu}(k)
= \int {d^D k \over (2\pi)^D} (D-1) k^2 D_T(k) 
{\partial \over \partial J} \Pi(k) .
\label{2ndEP}
\end{eqnarray}
Here we have used the fact that any gauge-invariant
regularization leads to the transverse form for the vacuum
polarization tensor of the gauge boson field:
\begin{eqnarray}
 \Pi_{\mu\nu}(k)  =  (g_{\mu\nu} k^2 - k_\mu k_\nu) \Pi(k).
\end{eqnarray}
The formulation of Thirring model as a gauge theory
justifies the use of gauge-invariant regularization and
guarantees the regularization-independence of the result.
\par
The above expression is slightly different from  the formula
for the chiral order parameter obtained by Ukita, Komachiya
and Fukuda \cite{UKF90}.  They have taken into
account up to the lowest order of the series expansion in
the coupling constant. In the gauged Thirring model the
corresponding term in the lowest order is given by 
\footnote{
The inversion up to the lowest order in the coupling
constant corresponds to the quenched approximation in the
following sense.  If we adopt the fermion propagator as the
nonlocal order parameter, the lowest-order inversion method
leads to the the SD equation in the quenched ladder
approximation where the photon propagator is
replaced with the bare one as well as the vertex
function, see Appendix.  }
\begin{eqnarray}
\int {d^D k \over (2\pi)^D} D^{(0)}_{\mu\nu}(k) 
{\partial \over \partial J} \Pi^{\mu\nu}(k)
= \int {d^D k \over (2\pi)^D}
 (D-1) k^2  D_T^{(0)}(k)
{\partial \over \partial J} \Pi(k),
\end{eqnarray}
where
\begin{eqnarray}
 D_T^{(0)}(k)  =  {1 \over \beta k^2+ M^2} ,
 \quad
 D_L^{(0)}(k)  =  {\xi \over k^2 + \xi M^2}.
\end{eqnarray}
\par
To obtain the explicit form of the vacuum polarization
function $\Pi(k)$, we here adopt the dimensional
regularization as a gauge-invariant regularization.  This
leads to the following well-known result:
\begin{eqnarray}
\Pi(k)  = - N_f {2\tr(1) \over (4\pi)^{D/2}} e^2 
\Gamma(2-{D \over 2})
\int_0^1 d\alpha {\alpha(1-\alpha) \over [J^2-k^2
\alpha(1-\alpha)]^{2-D/2}},
\label{vpD}
\end{eqnarray}
\begin{eqnarray}
{\partial \over \partial J} \Pi(k)  
= N_f {2\tr(1) \over (4\pi)^{D/2}} e^2 \Gamma(3-{D \over 2})
\int_0^1 d\alpha {2J \alpha(1-\alpha) \over [J^2-k^2
\alpha(1-\alpha)]^{3-D/2}}.
\label{dvpD}
\end{eqnarray}
For $D=4$, we get in the space-like region 
$k^2 = -k_E^2 < 0$
\begin{eqnarray}
 \Pi(k)  
=  N_f {e^2 \over 2\pi^2}  \left[ 
-{1 \over 6} \left( \ln {\Lambda^2 \over k_E^2} 
+ {5 \over 3} \right) + {J^2 \over k_E^2} 
+ {\cal O}(J^3) \right],
\label{vp4}
\end{eqnarray}
\begin{eqnarray}
{\partial \over \partial J} \Pi(k)  
=  N_f {e^2 \over \pi^2}  \left[ {J \over k_E^2} 
+ {2J^3 \over (k_E^2)^2} \ln {J^2 \over k_E^2}  
+ {\cal O}(J^5) \right],
\label{dvp4}
\end{eqnarray}
see \cite{IZ80,GR95} for details. 
The derivative Eq.~(\ref{dvp4}) is unchanged even if we
take other regularizations, e.g, the
Pauli-Villars regularization, see \cite{UKF90}, although
the finite constant in $\Pi(k)$ depends on the
regularization scheme adopted. 
\par

Now we move to the Euclidean space.  
In the lowest order, we have
\begin{eqnarray}
&& {1 \over 2N_f}
\int {d^4 k \over (2\pi)^4} D^{(0)}_{\mu\nu}(k) 
{\partial \over \partial J} \Pi^{\mu\nu}(k)
\nonumber\\
&=&   {3 \over 32\pi^4} \int_0^{\Lambda^2}  dx 
{x \over \beta x + M^2}
\left[ J 
+ {2J^3 \over x} \ln {J^2 \over x}  
+ {\cal O}(J^5) \right]
\nonumber\\
&=&   {\Lambda^2 \over 4\pi^2} 
{3 \over 8\pi^2} \Biggr[ \left( {1 \over \beta} 
- {1 \over \beta^2} {M^2 \over \Lambda^2} 
\ln {\beta \Lambda^2 +M^2 \over M^2} \right) J
\nonumber\\
&& + 2 {J^3 \over \Lambda^2} \int_0^{\Lambda^2}  dx 
{\ln {J^2 \over x} \over \beta x + M^2}
+ {\cal O}(J^5) \Biggr],
\end{eqnarray}
where we have defined $x := k_E^2=-k^2 >0$ and introduced
the ultraviolet (UV) cutoff $\Lambda$.
\par
Then the chiral order parameter is obtained as a function of
the external source (fermion mass):
\begin{eqnarray}
 \varphi := 4\pi^2 {\langle \bar \psi \psi \rangle \over
\Lambda^3} 
&=& \left\{ 1 + {3 \over 8\pi^2} 
\left[ {1 \over \beta}  - {\kappa \over \beta^2} 
\ln \left( 1+{\beta \over \kappa} \right) \right]
\right\} {J \over \Lambda}
\nonumber\\
&&+ {3 \over 4\pi^2}   {J^3 \over \Lambda^3}
\int_0^{\Lambda^2} dx {\ln J^2/x \over \beta x+M^2}
+ {\cal O}(J^5),
\label{opJ}
\end{eqnarray}
where we have defined the dimensionless coupling constant
$\kappa$ by 
\begin{eqnarray}
\kappa := {M^2 \over \Lambda^2} = {1 \over G_T\Lambda^2}.
\end{eqnarray}
Note that the above effective potential has no dependence
on the flavor number $N_f$ in the lowest order
\cite{KITE94}.

%\newpage
\section{Effective potential for chiral order parameter}
\setcounter{equation}{0}

Inverting Eq.~(\ref{opJ}) with respect to $J$ according
to the inversion formula (\cite{UKF90}, see Appendix),
we obtain
\begin{eqnarray}
  {J \over \Lambda} 
 =  \tau \varphi 
 - {3 \over 4\pi^2} \varphi^3 \int_0^1 dt {\ln
(\varphi^2/t)
\over \beta t + \kappa} + {\cal O}(\varphi^5),
\label{inverted}
\end{eqnarray}
where we have defined
\begin{eqnarray}
\tau = \tau(\beta,\kappa)
:= 1 - {3 \over 8\pi^2} {1 \over \beta}
\left[ 1  - {\kappa \over \beta} 
\ln \left( 1+{\beta \over \kappa} \right) \right] .
\end{eqnarray}
In order to obtain the effective potential, we use
the following relation (see Appendix A)  
between the effective potential and the
translation-invariant source $J$:
\begin{eqnarray}
 {\partial \tilde V(\varphi) \over \partial \varphi}
\equiv {J \over \Lambda} ,
\label{Legendre}
\end{eqnarray}
where we have defined the dimensionless effective
potential $\tilde V(\varphi)$ by
\begin{eqnarray}
\tilde V(\varphi) := 4\pi^2 {V(\varphi) \over \Lambda^4},
\quad
 \varphi := 4\pi^2 
 {\langle \bar \psi \psi \rangle \over \Lambda^3} .
\end{eqnarray}

\par
First, we consider the limit $\beta \rightarrow 0$. Then we
have
\begin{eqnarray}
 {J \over \Lambda} =   \tau(0,\kappa) \varphi 
 - {3 \over 4\pi^2} {1 \over \kappa} \varphi^3 
 \left( \ln \varphi^2  + 1 \right),
\end{eqnarray}
where 
\begin{eqnarray}
\tau(0,\kappa) = 1 -  {\kappa^c \over \kappa},
\quad
\kappa^c := {3 \over 16\pi^2}.
\end{eqnarray}
This reproduces the effective potential of the Thirring
model:
\begin{eqnarray}
 \tilde V(\varphi) =  {\tau \over 2} \varphi^2
 -  {\kappa_c \over \kappa} \varphi^4 
 \left( \ln \varphi^2 + {1 \over 2} \right) .
 \label{epTh4}
\end{eqnarray}
\par
Next, we consider the case of $\kappa \rightarrow 0$.  This
case corresponds to the pure QED$_4$.
Eq.~(\ref{inverted}) reproduces the result obtained in
ref. \cite{UKF90}.
\footnote{
Here we have used
$
\int^{\mu^2} dt {1 \over t} \ln {\varphi^2 \over t}
= - {1 \over 2} (\ln {\varphi^2 \over \mu^2})^2.
$
}
\begin{eqnarray}
 {J \over \Lambda} =   \tau(\beta,0) \varphi 
 +  {\beta^c \over \beta} \varphi^3 
 \left( \ln \varphi^2 \right)^2,
\end{eqnarray}
where 
\begin{eqnarray}
\tau(\beta,0) = 1 -  {\beta^c \over \beta},
\end{eqnarray}
\begin{eqnarray}
\beta^c := {3 \over 8\pi^2} = 2 \kappa^c.
\end{eqnarray}
Then the effective potential for QED$_4$ reads
\begin{eqnarray}
 \tilde V(\varphi) =  {\tau \over 2} \varphi^2
 +  {\beta^c \over \beta} \varphi^4 
 \left( \ln \varphi^2  \right)^2 .
 \label{epQED$_4$}
\end{eqnarray}

\par
For non-zero $\beta$ and $\kappa$, we have
\begin{eqnarray}
 {\partial \tilde V(\varphi) \over \partial \varphi}
\equiv {J \over \Lambda} 
 =    \tau \varphi 
 - 4{\kappa_c \over \kappa} \varphi^3 
 {\kappa \over \beta} \left\{
 \ln \left(1+{\beta \over \kappa}\right)  \ln \varphi^2  
 - {\rm PolyLog}\left[2,-{\beta \over \kappa}\right]
\right\}.
\end{eqnarray}
where we have introduced the polylogarithm function:
\begin{eqnarray}
 {\rm PolyLog}[2,z]
 := \int_z^0 {\log (1-t) \over t}dt
 = \sum_{n=1}^\infty {z^n \over n^2}.
\end{eqnarray}
Therefore, the effective potential for the chiral order
parameter in the gauged Thirring model is obtained for
arbitrary $\beta$ and $\kappa$:
\begin{eqnarray}
 \tilde V(\varphi) =   {\tau \over 2} \varphi^2
 + {\kappa_c \over \kappa} \varphi^4 
 {\kappa \over \beta} \left\{
 \ln \left(1+{\beta \over \kappa}\right) 
 \left( - \ln \varphi^2 + {1 \over 2} \right)
 + {\rm PolyLog}\left[2,-{\beta \over \kappa}\right]
\right\}.
\label{epgeneral}
\end{eqnarray}
This is an extension of the work \cite{UKF90} for QED$_4$.  It
should be noted that the
$\kappa=0$ case is somewhat special in the sense that the
$\varphi^4 (\ln \varphi^2)^2$ term does not appear if the
QED$_4$ limit is approached from non-zero $\kappa$, see
Eq.~(\ref{epgeneral}).   In fact, the $\kappa=0$ case has an
ambiguity coming from the lower bound of the integral
Eq.~(\ref{inverted}).
\par
The order parameter $\varphi$ can have a nontrivial minimum
for $\tau<0$, as long as the second $\varphi^4$ part takes 
the positive value and convex in $\varphi$ when $\tau = 0$.
This is satisfied for the effective potentials
Eq.~(\ref{epQED$_4$}) and Eq.~(\ref{epTh4}) at least in the
small field region $\varphi \ll 1$, although
Eq.~(\ref{epTh4}) is negative for the large field $\varphi >
{\cal O}(1)$, Fig.~2 (a), (b).
\par
The term 
$\varphi^4 \left(- \ln \varphi^2 + {1 \over 2} \right)$
is positive for $\varphi \ll 1$.
However,  ${\rm PolyLog}[2,-t]$ is
negative for $t:=\beta/\kappa>0$ and the absolute value is
monotone increasing in $t$.  
Nevertheless, there is a certain value $\varphi_0>0$ such
that the second term in (\ref{epgeneral}) is positive for
$0 < \varphi < \varphi_0$, no matter how large the ratio
$\beta/\kappa$ may be, see Fig.~2 (c).
\par
Note that the effective potential (\ref{epgeneral})
obtained in the lowest order does not have any $N_f$
dependence.  In the original work \cite{UKF90}, the
effective potential (\ref{epQED$_4$}) has been used to
estimate the critical coupling of quenched QED$_4$. However,
it is well known that the effective potential (\ref{epTh4})
is justified in the limit
$N_f \rightarrow \infty$, see e.g. \cite{Miransky93}. 
Moreover, the critical point $\beta^c$ of QED$_4$ obtained from
the effective potential (\ref{epQED$_4$}) agrees with the
result for $N_f > 1$ obtained from the SD equation
\cite{KITE94}.
In view of these, therefore, it is reasonable to
consider that our result (\ref{epgeneral}) will be applied
to the case of relatively large $N_f$ in the gauged
Thirring model.

%\newpage
\section{Phase diagram and critical behavior}
\setcounter{equation}{0}

\subsection{critical line}
We can specify the location of the critical line by putting
$\tau(\beta,G_T)=0$ which separates the 
spontaneous chiral-symmetry-breaking phase from the  
chiral-symmetric phase.
If we define the reduced coupling constants:
\begin{eqnarray}
 \tilde \beta := {\beta \over \beta_c} ,
 \quad
 \tilde \kappa := {\kappa \over \kappa_c} ,
\end{eqnarray}
by using the critical couplings $\beta_c$ of pure QED  and
 $\kappa_c$ of Thirring model,
the equation of the critical line is written as
\begin{eqnarray}
 \tilde \beta  
 + {1 \over 2} {\tilde \kappa \over \tilde \beta} \ln
(1 + 2{\tilde \beta \over \tilde \kappa})   = 1  .
  \label{rcl}
\end{eqnarray}
In the phase diagram $(\beta, \kappa)$ 
the chiral symmetry is spontaneously broken below this
critical line.
The critical line Eq.~(\ref{rcl}) is plotted in
Fig.~3. 
It is easy to show that the critical line reduces  to 
\begin{eqnarray}
\tilde \kappa = 1 - {4 \over 3} \tilde \beta
+ {\cal O}(\beta^2), 
\end{eqnarray}
in the neighborhood of $(0,\kappa_c)$.
\footnote{
This shows good agreement with the result of Monte
Carlo simulation of the corresponding model of compact
version on the lattice by Jersak et al.
\cite{FJ95}.
}
\par
The critical line (\ref{rcl}) obtained from the chiral
order parameter is gauge independent, although we adopted
the special gauge to simplify the calculation. 
Of course, it should be noted that the other
critical lines can exist, if we use other
order parameters (local or nonlocal). Hence the critical
line (\ref{rcl}) is not the only one for this model. 
Actually, by taking into account the
order parameter containing the scalar field, more rich phase
structure will be probed, see
\cite{FJ95}.
\par

\subsection{critical exponent and order of the transition}

The effective potential (\ref{epgeneral}) shows that the
chiral order parameter exhibits the power law behavior in
the neighborhood of the critical line, apart from the
logarithmic correction:
\begin{eqnarray}
 \langle \bar \psi \psi \rangle \sim 
 |\tau|^{\beta_{ch}} |\log |\tau| |^a,
\end{eqnarray}
where $\tau$ denotes the deviation from the critical
line in the two-parameter space $(\beta,\kappa)$.
Our result is consistent with the mean-field theory
and the critical exponent $\beta_{ch}$ for the chiral order
parameter takes the mean-field value $\beta_{ch}=1/2$,
irrespective of the values of the coupling constant.
The effective potential for the chiral order parameter
Eq.~(\ref{epgeneral}) shows that the chiral phase transition
is of the second order on the whole critical line.
A possible explanation of this disagreement with the
lattice result for the compact version \cite{FJ95} will
be given in the final section.

\newpage
\section{Set up of the Schwinger-Dyson equation}
\setcounter{equation}{0}

In the following sections, we study the phase structure
in detail by the use of the Schwinger-Dyson (SD)
equation.  The effective potential for the chiral order
parameter obtained in the previous section is gauge
independent.  However, as is well known, the solution of
the SD equation for the fermion propagator is never gauge
independent and depends on the chosen gauge.  
However, the criterion (the renormalized coupling constant
or the mass ratio) which we use in section 11 for the
nontriviality of the gauged Thirring model is gauge
independent and does not depend on the specific gauge.
Therefore, we can hope that the final result obtained from
the SD equation on the gauge-invariant quantities does not
depend on the choice of gauge which is necessary to
actually solve the SD equation.

\subsection{introduction of the nonlocal $R_\xi$ gauge}

We can write down the SD equation for the
gauged Thirring model.   
The full fermion propagator is written as
\begin{equation}
  S(p) = [A(p^2) \gamma^\mu p_\mu  - B(p^2)]^{-1},
\end{equation}
in accord with the bare fermion propagator 
\begin{equation}
  S_0(p) = (\gamma^\mu p_\mu - m_0)^{-1}.
\end{equation}
For the gauged Thirring model with the Lagrangian
(\ref{abeliangTh}), the exact SD equation for the full
fermion propagator is given by
\begin{equation}
  S^{-1}(p) = S_0^{-1}(p) + \int {d^Dq \over (2\pi)^D}
  \gamma_\mu S(q) \Gamma_\nu(p,q) D_{\mu\nu}(p-q),
  \label{SD}
\end{equation}
where $\Gamma_\nu(p,q)$ is the full vertex function 
and $D_{\mu\nu}(p-q)$ is the full gauge boson propagator.
Note that this SD equation can be decomposed into a pair
of integral equations for the wavefunction
renormalization function $A(p^2)$ and the mass function
$B(p^2)$.
\par
The SD equation for the full fermion propagator constitutes
the closed set of equations together with the SD equations
for the full vertex function and the full gauge boson
propagator.
In order to solve the SD equation for the fermion
propagator, therefore, we must anyway specify the full
vertex function
$\Gamma_\nu(p,q)$ and the full gauge boson propagator
$D_{\mu\nu}(p-q)$.
In most cases, an ansatz for the vertex function is
adopted instead of solving the SD equation for the
vertex function, although it is in principle possible to
solve all the SD equation simultaneously.   The simplest
choice is the bare vertex approximation:
\begin{equation}
 \Gamma_\mu(p,q)  \equiv  \gamma_\mu. 
\end{equation}
This leads to the simplest ladder approximation.
For this approximation to be consistent with the
Ward-Takahashi identity which is a consequence of the gauge
invariance of the theory, 
\footnote{
In order for the vertex function to be consistent with the
Ward-Takahashi identity, the ansatz for the vertex
function should include $S$, i.e. $A$ and $B$.
However, this requirement is not sufficient to determine
the vertex uniquely, see ref.
\cite{Kondo92} and references therein. 
} 
there should be no wavefunction renormalization for the
fermion.  Therefore, the bare vertex approximation should
self-consistently yield the result: 
\begin{equation}
 A(p^2) \equiv 1, 
 \label{nowfr}
\end{equation}
as the solution of SD equation Eq.~(\ref{SD}).  
This requirement is satisfied if we introduce the
momentum-dependent function $\eta(k^2)$ in the
gauge boson propagator as
\begin{equation}
  D_{\mu\nu}(k) 
  = D_T(k)
    \left[ g_{\mu\nu} - \eta(k^2) \frac{k_\mu k_\nu}{k^2}
\right] .
\label{gbpropanl}
\end{equation}
As is shown in the next subsection, the function
$\eta(k^2)$ can be chosen so that the SD equation
Eq.~(\ref{SD}) in the bare vertex approximation leads to
the solution Eq.(\ref{nowfr}) for the wavefunction
renormalization $A(p^2)$ \cite{GSC90,KM92}.
Therefore, in this setup, we have only to solve the SD
equation for the fermion mass function $B(p^2)$ in the
nonlocal gauge.
\par
The introduction of momentum-dependent function
$\eta(k^2)$ in the full gauge boson propagator corresponds
to introducing the nonlocal gauge fixing term into the total
Lagrangian instead of the usual one, see
Eq.~(\ref{nlgfterm}).  
By comparing (\ref{gbpropa}) with (\ref{gbpropanl}),
the correspondence between $\xi$ and
$\eta$ is given in momentum space as follows.
\begin{eqnarray}
  \xi(k^2) = {[1-\eta(k^2)]k^2 \over 
\eta(k^2) M^2 + \beta k^2 -\Pi(k)},
  \quad
  \eta(k^2) = {k^2-\xi(k^2)[\beta k^2 - \Pi(k)] \over 
k^2 + \xi(k^2) M^2}.
\end{eqnarray} 

\subsection{SD equation  in the nonlocal $R_\xi$ gauge}
\par
The SD equation is decomposed into a pair of integral
equations:
\begin{eqnarray}
 A(p^2) &=& 1 
 + {\tr[\Sigma(p) \gamma^\mu p_\mu] \over p^2 \tr(1)},
\nonumber\\ 
 B(p^2) &=& m_0 
 + {\tr[\Sigma(p)] \over p^2 \tr(1)},
 \label{decompSD}
\end{eqnarray} 
where $\Sigma$ is the self-energy part:
\begin{equation}
  \Sigma(p) := \int {d^Dq \over (2\pi)^D}
  \gamma_\mu S(q) \gamma_\nu D_{\mu\nu}(p-q).
  \label{selfenergy}
\end{equation}
The SD equation for the fermion wave function
renormalization $A$ is given by
\begin{eqnarray}
 && p^2 A(x) -  p^2
\nonumber\\ &=& e^2 \int {d^Dq \over (2\pi)^D} 
{A(q^2) \over q^2 A^2(q^2)+B^2(q^2)}  
\nonumber\\ &&  \times k^2 D_T(k)
\left[ (D-2) {p \cdot q \over k^2} + \left( {p
\cdot q \over k^2} - 2 {p^2q^2-(p \cdot q)^2 \over
k^4} \right) \eta(k^2) \right] .
\end{eqnarray} 
On the other hand, the SD equation for the fermion
mass function $B$ reads
\begin{eqnarray}
 B(p^2) &=& m_0 + e^2 \int {d^Dq \over (2\pi)^D} {B(q^2)
\over q^2 A^2(q^2)+B^2(q^2)}
D_T(k) [D-\eta(k^2)] .
\label{SDB}
\end{eqnarray} 
\par
Separating the angle $\vartheta$ defined by
\begin{equation}
  k^2 := (q-p)^2 = x + y -2\sqrt{xy} \cos\vartheta,
\  x := p^2, \qquad y := q^2,
\end{equation}
we find
\begin{eqnarray}
 && x A(x) -  x
\nonumber\\ &=&   C_D e^2 \int_0^{\Lambda^2} dy 
{y^{(D-2)/2}A(y) \over y A^2(y)+B^2(y)} 
  \int_0^{\pi} d \vartheta \sin ^{D-2} \vartheta   
  D_T(k)
\nonumber\\ && \times k^2  \left\{  {\sqrt{xy}
\cos \vartheta [D-2+\eta(k^2)] \over k^2}
 - 2 {xy-(\sqrt{xy} \cos \vartheta)^2 \over k^4}  
\eta(k^2) \right\},
\end{eqnarray} 
and
\begin{eqnarray}
 B(x) = m_0 + C_D e^2  \int_0^{\Lambda^2} dy 
 {y^{(D-2)/2}B(y) \over y A^2(y)+B^2(y)} K(x,y),
 \label{SDeqB}
\end{eqnarray} 
\begin{eqnarray}
K(x,y) := \int_0^{\pi} d \vartheta \sin ^{D-2}
\vartheta   D_T(k) [D-\eta(k^2)],
\label{kernel}
\end{eqnarray} 
where
\begin{equation}
 C_D := {1 \over 2^D \pi^{(D+1)/2} \Gamma({D-1
\over 2})} .
\end{equation}

\par
Following the same procedure as given in the
Appendix of ref.\cite{KEIT95}, it turns out that
 the requirement $A(p^2) \equiv 1$
is achieved if  $\eta(k^2)$ satisfies the following
equation:
\begin{eqnarray}
 (z^{D-1} D_T(z) \eta(z))' + (D-2) z^{D-1} D_T'(z) 
 \equiv 0,
 \quad z := k^2.
\end{eqnarray}
This is simply solved as follows.
\begin{eqnarray}
 \eta(z) &=& - {D-2 \over z^{D-1} D_T(z)} \int_0^z dt
D_T'(t) t^{D-1} ,
\label{nlg1}
\end{eqnarray} 
where we have assumed that 
$[z^{D-1} D_T(z) \eta(z)]|_{z=0}=0$ so as to
eliminate the $1/z^{D-1}$ singularity in $\eta(z)$. 
This should be checked after having obtained the function
$\eta(z)$.
Alternatively, we can write
\begin{eqnarray}
 \eta(z) =  (D-2) \left[
{D-1 \over z^{D-1} D_T(z)} \int_0^z dt D_T(t) t^{D-2} -1
\right],
\label{nlg2}
\end{eqnarray} 
where we have assumed that 
$[z^{D-1} D_T(z)]|_{z=0}=0$.
\par
Thus, once the function $D_T(k^2)$ is given, we can find the
nonlocal gauge $\eta(k^2)$ according to the above
formulae, Eq.~(\ref{nlg1}) or Eq.~(\ref{nlg2}), so that
$A(k^2) \equiv 1$ follows in agreement with the bare vertex
approximation.  Then we have only to solve
Eq.~(\ref{SDB}) for the fermion mass function $B(p^2)$.
\begin{eqnarray}
 B(p^2) &=& m_0 + e^2 \int {d^Dq \over (2\pi)^D} {B(q^2)
\over q^2 +B^2(q^2)}D_T(k) [D-\eta(k^2)] .
\label{SDBnlg}
\end{eqnarray} 
Note that, if we use the following definition of gauge
boson propagator:
\begin{equation}
  D_{\mu\nu}(k) 
= D_T(k)\left(g_{\mu\nu} - {k_\mu k_\nu \over k^2}\right)
+ D_L(k) {k_\mu k_\nu \over k^2},
\end{equation}
the integrand is replaced as follows.
\begin{equation}
D_T(k) [D-\eta(k^2)] = (D-1)D_T(k)+D_L(k) .
\end{equation}

%\newpage
\section{CJT effective action in the nonlocal gauge}
\setcounter{equation}{0}

\subsection{CJT effective potential}
Now we derive the Cornwall-Jackiw-Tomboulis (CJT) action 
\cite{CJT74} such that the stationary point of the CJT
effective action as a functional of the fermion
propagator gives the SD equation for the fermion propagator
in the nonlocal gauge.  In the case of QED$_3$, it has been
already given in ref.~\cite{KM95}.
\par
The CJT effective action \cite{CJT74} is given by
\begin{eqnarray}
 \Gamma[S] &=& \Gamma_0[S] + \Gamma_1[S],
\\ && 
 \Gamma_0[S] := - \Tr[\ln (S^{-1}S_0) + S_0^{-1}S - 1],
\\&& 
\Gamma_1[S] := {1 \over 2} 
\Tr[\gamma^\mu S \gamma^\nu S D_{\mu\nu}] ,
\end{eqnarray} 
where we have chosen $\Gamma_0[S]$ so that
$\Gamma_0[S=S_0]=0$.
Indeed, this leads to the SD equation at the stationary
point (see Fig.~4):
\begin{eqnarray}
0 = {\delta  \Gamma[S] \over \delta S}
=  S^{-1} - S_0^{-1} +
\gamma^\mu S \gamma^\nu D_{\mu\nu} .
\end{eqnarray} 

\par
Explicit evaluations in momentum space lead to
\begin{eqnarray}
 {\Tr[\ln (S^{-1}S_0)] \over \Omega}
= {\tr(1) \over 2} \int {d^Dp \over (2\pi)^D} 
\ln \left[ 1 + {B^2(p^2) \over p^2} \right],
\end{eqnarray} 
\begin{eqnarray}
 {\Tr[S_0^{-1}S] \over \Omega}
 =   \tr(1) \int {d^Dp \over (2\pi)^D} 
  {B(p^2)[m_0-B(p^2)] \over p^2+B^2(p^2)} ,
\end{eqnarray} 
where $\Omega$ is the volume of spacetime,
$\Omega = \int d^D x$ and $\tr(1)$ denotes the trace over
the spinor indices, e.g. $\tr(1)=4$ for $D=4$. The second
part is also evaluated as
\begin{eqnarray}
{\Gamma_1[S]  \over \Omega} 
&=& {1 \over 2} \int {d^Dp \over (2\pi)^D}
  \int {d^Dq \over (2\pi)^D}
  \Tr[\gamma^\mu S(p) \gamma^\nu S(q)] D_{\mu\nu}(p-q) ,
\nonumber\\
&=& {\tr(1) \over 2} \int {d^Dp \over (2\pi)^D}
  \int {d^Dq \over (2\pi)^D} 
  {[D-\eta(k^2)]D_T(k) B(p^2)B(q^2) \over 
  [p^2+B^2(p^2)][q^2+B^2(q^2)]},
\end{eqnarray} 
where we have used the observation that the $A$-dependent
term appearing in the integrand vanishes if
we adopt the nonlocal gauge obtained in the previous
section, see Eq.~(\ref{decompSD}).
\par
For the translation invariant $S$, i.e., $S(x,y)=S(x-y)$,
we can define the CJT effective potential
$V$ as
\begin{eqnarray}
  V[B] := - {\Gamma[S] \over \Omega} ,
\end{eqnarray} 
\begin{eqnarray}
 V[B] &=& - {\tr(1) \over 2} \int {d^Dp \over (2\pi)^D}
\left\{ 
 \ln \left[ 1 + {B^2(p^2) \over p^2} \right] 
 - {2B(p^2)[B(p^2)-m_0] \over p^2+B^2(p^2)} \right\}
\nonumber\\
&-& {\tr(1) \over 2} \int {d^Dp \over (2\pi)^D}
  \int {d^Dq \over (2\pi)^D} 
  {D_T(k) [D-\eta(k^2)]B(p^2)B(q^2) \over 
  [p^2+B^2(p^2)][q^2+B^2(q^2)]}.
  \label{EPgTh}
\end{eqnarray} 
\par
If we separate the integration variable into the radial
part $x:=p^2$ and the angular part $\vartheta$, and
introduce the UV cutoff for the radial part, the effective
potential reads
\begin{eqnarray}
 V[B] &=& - {\tr(1) \over 2} 
 \tilde C_D \int_0^{\Lambda^2} dx x^{(D-2)/2}
\left\{ 
 \ln \left[ 1 + {B^2(x) \over x} \right] 
 - {2B(x)[B(x)-m_0] \over x+B^2(x)} \right\}
\nonumber\\
&-& {\tr(1) \over 2} \tilde C_D C_D \int_0^{\Lambda^2} dx 
\int_0^{\Lambda^2} dy 
  {x^{(D-2)/2}B(x) y^{(D-2)/2}B(y) \over 
  [x+B^2(x)][y+B^2(y)]} K(x,y).
\label{EPgThcutoff}
\end{eqnarray} 
where $K(x,y)$ is the kernel defined by Eq.~(\ref{kernel})
and we have defined
\begin{eqnarray}
  \tilde C_D  := C_D 
  {\Gamma({1 \over 2}) \Gamma({D-1 \over 2}) 
  \over \Gamma({D \over 2})}
  = {1 \over 2^D \pi^{D/2} \Gamma({D \over 2})} .
\end{eqnarray} 
\par
Indeed, the variation of the CJT effective potential
with respect to $B$, i.e.,
\begin{eqnarray}
  {\delta \over \delta B(p^2)} V[B] = 0
\end{eqnarray} 
yields the SD equation Eq.~(\ref{SDBnlg}).
However, it should be noted that this is possible only when
the functions $D_T$ and $\eta$ do not include any
$B$-dependence.   Hence the following argument is
restricted to the quenched case: $\Pi \equiv 0$ or the
restricted (approximated) unquenched case
\cite{Kondo91a,Kondo91b} where the massless fermion is used
in evaluating the vacuum polarization function.
\par
If we denote the solution of the SD equation by $B_{sol}$,
the effective potential at the stationary point reads
\begin{eqnarray}
 V[B_{sol}]  =  
- {\tr(1) \over 2} \int {d^Dp \over (2\pi)^D}
 \left\{ \ln \left[ 1 + {B_{sol}^2(p^2) \over p^2} \right]
 - {B_{sol}(p^2)[B_{sol}(p^2)-m_0] \over p^2+B_{sol}^2(p^2)}
 \right\} .
 \label{epstationay}
\end{eqnarray} 
Note that the SD equation has always a trivial solution
when $m_0=0$.
This expression shows that the nontrivial solution
$B_{sol}\not= 0$ gives lower effective potential than
the trivial one $B_{sol} \equiv 0$, i.e.,
$V[B_{sol}] < V[B_{sol} \equiv 0]=0$.
This is shown by rewriting Eq.~(\ref{epstationay}) as
\begin{eqnarray}
 V[B_{sol}] &=& - {\tr(1) \over 2} \tilde C_D
\int_0^{\Lambda^2}  dp^2  
 (p^2)^{(D-2)/2} g \left({B_{sol}^2(p^2) \over p^2}\right),
 \nonumber\\
 g(t) &=& \ln(1+t) - {t \over 1+t},
 \label{epstationary2}
\end{eqnarray} 
since $g(0)=0$ and $g(t)$ is positive and
monotone increasing function of $t$.
Therefore, if the SD equation has a nontrivial
solution in the chiral limit $m_0=0$, the
 chiral-symmetry-breaking solution should be realized.

\subsection{scale invariance}
By introducing the scaling parameter $\mu$, we can rewrite
the effective potential (\ref{EPgThcutoff}) as
\begin{eqnarray}
 &&- {V[B] \over  {\tr(1) \over 2} \tilde C_D}  
\nonumber\\
 &=&  
 \mu^{-D} \int_0^{\mu^2 \Lambda^2} d x x^{(D-2)/2}
\left\{ 
 \ln \left[ 1 + {B_\mu^2(x) \over x} \right] 
 - {2B_\mu(x)[B_\mu(x)-\mu m_0] \over  x+B_\mu^2(x)}
\right\}
\nonumber\\
&+& \mu^{-2D+2} C_D \int_0^{\mu^2 \Lambda^2} dx 
\int_0^{\mu^2 \Lambda^2} dy 
K \left({x \over \mu^2},{y \over \mu^2}\right)
{x^{{D-2 \over 2}}B_\mu(x) 
y^{{D-2 \over 2}}B_\mu(y)
\over [x+B_\mu^2(x)][y+B_\mu^2(y)]} ,
\end{eqnarray} 
where we have defined the scaled mass function $B_\mu(p^2)$
by
\begin{eqnarray}
 B_\mu(p^2) := \mu B \left({p^2 \over \mu^2}\right) .
\end{eqnarray} 
We denote the effective potential (\ref{EPgThcutoff}) by
$V[B;\Lambda,m_0; \{\alpha_i\}]$ where $\{\alpha_i\}$
denotes the set of all the other bare parameters appearing
in the kernel
$K(x,y;\{\alpha_i\})$.
Then the effective potential (\ref{EPgThcutoff}) satisfies
\begin{eqnarray}
V[B;\Lambda,m_0; \{\alpha_i\}] = \mu^{-D} 
V \left[B_\mu; \mu \Lambda , \mu m_0; \{\alpha_i'\} \right],
\label{scalingEP}
\end{eqnarray} 
or equivalently
\begin{eqnarray}
V[B_\mu;\Lambda,m_0; \{\alpha_i'\}] = \mu^{D} 
V \left[B;{\Lambda \over \mu}, {m_0 \over \mu};
\{\alpha_i\} \right],
\label{scaleinv}
\end{eqnarray} 
if the kernel satisfies the following property
\begin{eqnarray}
 K \left( {p^2 \over \mu^2}, {q^2 \over \mu^2};
\{\alpha_i\}
\right)
 = \mu^{D-2} K(p^2, q^2; \{\alpha_i'\}),
 \label{scalingkernel}
\end{eqnarray} 
under the appropriate transformation between the new bare
parameters ${\alpha_i'}$ and the original bare parameters
${\alpha_i}$, i.e. $\alpha_i' = f_i(\{\alpha_j\}_j, \mu)$.
\par
If we require (\ref{scalingkernel}) for the kernel, the
scaled mass function satsifies the same type of the SD
equation as Eq.~(\ref{SDeqB}) for $B(p^2)$ with the bare
parameters
$m_0$ and
$\{\alpha_i\}$:
\begin{eqnarray}
 B_\mu(x) = \mu m_0  + C_D \int_0^{\mu^2\Lambda^2} dy 
 {y^{(D-2)/2}B_\mu(y) \over y+B_\mu^2(y)} 
 K(x,y; \{\alpha_i'\}),
\end{eqnarray} 
but in the scaled range of momentum
$p^2 \in [0, (\mu\Lambda)^2]$ and with the transformed bare
parameters, 
$\mu m_0$ and $\{\alpha_i'\}$.
\par
In the quenched QED$_4$, the relation 
(\ref{scalingkernel}) is trivially satisfied without any
change of the coupling constant, since 
the kernel has the form:
\begin{eqnarray}
 K(x,y) = K(x,y; e)
\propto e^2 \left[{\theta(x-y) \over x} 
 + {\theta(y-x) \over y} \right].
\end{eqnarray} 
In the gauged NJL$_4$ model with the fixed gauge coupling
\cite{KTY93}, the kernel is given by
\begin{eqnarray}
 K(x,y) = K(x,y; e, G)
\propto e^2 \left[{\theta(x-y) \over x} 
 + {\theta(y-x) \over y} \right] + G.
\end{eqnarray} 
Hence the relation (\ref{scalingkernel}) is
satisfied when 
\begin{eqnarray}
e' = e,
\quad
G' = \mu^{-2} G,
\end{eqnarray} 
i.e. the gauge coupling constant
is unchanged and only the four-fermion coupling constant
$G$ is scaled. 
\par
For the gauged Thirring model, such a transformation
between the bare parameters is given by
\begin{eqnarray}
\beta ' = \beta,
\quad
G_T' = \mu^{-2} G_T,
\end{eqnarray} 
as can be seen in the next section, although the
explicit kernel is different from the gauged NJL model.
However, it should be noted that this does not give the
exact RG flow in the gauged Thirring model, see section 11.

\par
When all the bare parameters are unchanged,
differentiating both sides of Eq.~(\ref{scaleinv}) with
respect to $\mu$ and putting
$\mu=1$, we obtain
\begin{eqnarray}
  V \left[B_{sol}; \Lambda ,  m_0 \right]
  = {1 \over D} \left[ 
  2 \Lambda^2 {\partial \over \partial \Lambda^2} 
  + m_0 {\partial \over \partial m_0} \right]
  V \left[B; \Lambda ,  m_0 \right] \Big|_{B=B_{sol}}.
\end{eqnarray} 
where we have used
\begin{eqnarray}
{\delta V \over \delta B} \Big|_{B=B_{sol}}=0.
\end{eqnarray} 
\par
In applying this formula to Eq.~(\ref{epstationary2}),
note that 
\begin{eqnarray}
&& \left[ 
  2 \Lambda^2 {\partial \over \partial \Lambda^2} 
  + m_0 {\partial \over \partial m_0} \right]
  V \left[B; \Lambda ,  m_0 \right] \Big|_{B=B_{sol}}
  \nonumber\\&&
  \not= 
\left[ 
  2 \Lambda^2 {\partial \over \partial \Lambda^2} 
  + m_0 {\partial \over \partial m_0} \right]
  V \left[B_{sol}; \Lambda ,  m_0 \right] .
\end{eqnarray} 
Hence, it is wrong that 
\begin{eqnarray}
  V \left[B_{sol}; \Lambda ,  m_0 = 0 \right]
  = -  \tr(1)   {\tilde C_D \over D} 
 \Lambda^D g \left({B^2(\Lambda^2) \over \Lambda^2}\right),
\end{eqnarray} 
\begin{eqnarray}
 g(t) = \ln(1+t) - {t \over 1+t}
 = 
{{{t^2}}\over 2} - {{2\,{t^3}}\over 3} + 
  {{3\,{t^4}}\over 4} + {{{\rm O}(t)}^5}.
\end{eqnarray}

\newpage
\section{Solution of SD equation}
\setcounter{equation}{0}

In the following sections we adopt the quenched (ladder)
approximation in which the vacuum polarization function
$\Pi$ is neglected in $D_T$:  
\begin{eqnarray}
 D_T(k^2) \rightarrow
 D_T^{(0)}(k^2) := {1 \over \beta k^2 + M^2}.
\end{eqnarray} 
This approximation corresponds to the $N_f
\rightarrow 0$ limit.
Hence the following result will hold for the relatively
small $N_f$, in sharp contrast with the result of the
effective potential.
In the quenched limit, the nonlocal gauge is given by
\begin{eqnarray}
&& \eta(k^2) 
\nonumber\\
&=&   - 2 +
{6 \over (k^2)^{3} D_T^{(0)}(k^2)} \int_0^{k^2} dt
D_T^{(0)}(t) t^{2} 
\nonumber\\
&=& - 2 +
{6(\beta k^2+M^2) \over (k^2)^{3}} \int_0^{k^2} dt
{t^{2} \over \beta t + M^2}  
\nonumber\\
&=& - 2 +
 3 \left(1+{M^2 \over \beta k^2} \right) 
\left[ 1 - 2 {M^2 \over \beta k^2} 
+ 2 {M^4 \over \beta^2 k^4} \ln \left( 1+{\beta k^2 \over
M^2} \right) \right].
\label{nlg3+1}
\end{eqnarray} 
\par
First we observe two limits:
In the pure QED limit: $M \rightarrow 0$, $\eta(k^2)$
approaches a constant uniformly in $k^2$:
\begin{eqnarray}
 \eta(k^2) \rightarrow 1,
\end{eqnarray} 
which corresponds to the Landau gauge, $\xi = 0$.
\par
In the pure Thirring-model limit: 
$\beta \rightarrow 0$, on the other hand, $\eta(k^2)$
vanishes:
\begin{eqnarray}
 \eta(k^2) \rightarrow 0.
\end{eqnarray} 
This corresponds to the unitary gauge,
$\xi \rightarrow \infty$.
The nonlocal gauge is plotted as a function of 
$\beta k^2/M^2$ in Fig.~5.
The function $\eta(k^2)$ increases monotonically and has
values between 0 and 1.
\par
In the nonlocal gauge~(\ref{nlg3+1}), we
study the solution of the SD equation:
\begin{eqnarray}
 B(x) = m_0 + {1 \over 16\pi^2}
 \int_0^{\Lambda^2}dy {yB(y) \over y+B^2(y)} \tilde K(x,y),
 \label{SDB3+1}
\end{eqnarray} 
where the kernel $\tilde K(x,y)$ is given by
\begin{eqnarray}
&&  \tilde K(x,y) :=
{2 \over \pi}
\int_0^\pi d\vartheta \sin^2 \vartheta D_T(k)[4-\eta(k^2)],
\label{kernel4}
\end{eqnarray} 
and we have defined the squared Euclidean momenta:
\begin{eqnarray}
  x := p^2 > 0, \quad y := q^2 > 0.
\end{eqnarray} 
Here we have introduced the UV cutoff $\Lambda$ as an
upper bound of integration.
\par
In the chiral limit $m_0=0$, the equation
Eq.~(\ref{SDB3+1}) has a trivial solution $B(p^2) \equiv 0$
irrespective of values of the coupling constants.   
If the SD equation has a nontrivial solution in the
chiral limit $m_0=0$,  then the fermion mass is dynamically
generated and the chiral symmetry is spontaneously
broken, since the nontrivial solution gives lower
effective potential than the trivial solution.  Actually it
has been shown
\cite{MN74} that this phenomenon occurs in QED$_4$, if the
gauge coupling is allowed to be strong. 
This suggests the existence of the critical point
$\beta_e^c$ such that the nontrivial solution exists
only for 
$\beta_e := 4\pi^2/e^2 < \beta_e^c$
(supercritical or strong coupling region). 
\par
The simplest way for obtaining the value of the critical
coupling and the explicit form of the nontrivial solution is
to look for the bifurcation solution of the nonlinear
integral equation \cite{Atkinson,Kondop}.
The bifurcation solution from the trivial one, if any, is
obtained from the following 
\footnote{Here we have assumed that the kernel
$\tilde K(x,y)$ does not include the function $B(x)$.
This is satisfied at least in the quenched ladder
approximation.} 
linearized equation \cite{Tricomi} of Eq.~(\ref{SDB3+1}):
\begin{eqnarray}
 b(x) =   {1 \over 16\pi^2}
 \int_0^{\Lambda^2}dy 
 {\partial \over \partial B(y)} \left[
 {yB(y) \over y+B^2(y)} \right] \Biggr|_{B=0} 
 \tilde K(x,y) b(y),
\label{bif0SDB3+1}
\end{eqnarray} 
namely,
\begin{eqnarray}
 b(x) =   {1 \over 16\pi^2}
 \int_0^{\Lambda^2}dy  
 \tilde K(x,y)  b(y).
\label{bifSDB3+1}
\end{eqnarray} 
As the linearized equation does not determine the scale of
the solution, we impose the following condition to normalize
the solution: $b(p^2=m^2)=m$. 
The nontrivial solution corresponds to
the bifurcation solution of the original nonlinear integral
equation. The critical coupling constant $\beta_e^c$ is
obtained as the bifurcation point of the solution, at
which  the number of the solution changes. 
\par
Another way of linearizing Eq.~(\ref{SDB3+1}) is
to study the following equation \cite{Miransky85}:
\begin{eqnarray}
 B(x) = m_0 + {1 \over 16\pi^2}
 \int_0^{\Lambda^2}dy {yB(y) \over y+B^2(0)} \tilde K(x,y),
\label{MirSDB3+1}
\end{eqnarray} 
where the normalization of the solution is given by
$m=B(0)$.
\par
Both method give the same result at least on the critical
point, if they have the nontrivial solution:
$\beta_e^c = 3$ for quenched QED$_4$.

\subsection{strong gauge-coupling region: $\beta \ll 1$}
\par
First, we consider the strong gauge-coupling region, i.e.,
$\beta \ll 1$.  In this region, the nonlocal
gauge~(\ref{nlg3+1}) can be expanded as follows:
\begin{eqnarray}
 \eta(k^2) &=&  
 \sum_{m=1}^{\infty} {6 (-1)^{m-1} \over (m+2)(m+3)} 
 \left({\beta k^2 \over M^2} \right)^m
 \nonumber\\
 &=&  {1 \over 2}{\beta k^2 \over M^2} 
 - {3 \over 10}\left({\beta k^2 \over M^2} \right)^2
 + {\cal O}(\beta^3).
\end{eqnarray} 
In accord with this expansion, we have
\begin{eqnarray}
 D_T^{(0)}(k) 
 = {1 \over M^2} \sum_{m=0}^{\infty} (-1)^m 
 \left({\beta k^2 \over M^2} \right)^m .
\end{eqnarray} 
Therefore the integrand of the kernel of the integral
equation is given by
\begin{eqnarray}
 D_T^{(0)}(k)[4-\eta(k^2)]
&=& {6 \over M^2} 
 \sum_{m=0}^{\infty} {m+2 \over m+3}(-1)^m 
 \left({\beta k^2 \over M^2} \right)^m
 \nonumber\\
 &=&   {4 \over M^2} \left[ 1 
 - {9 \over 8} {\beta k^2 \over M^2}  
 + {6 \over 5}\left({\beta k^2 \over M^2} \right)^2 \right]
 + {\cal O}(\beta^3).
\end{eqnarray} 
Making use of the integration formulae,
\begin{eqnarray}
 \int_0^\pi d\vartheta \sin^2 \vartheta (\cos \vartheta)^n
 = \cases{ {1 \over 2}\pi &(n = 0) \cr
  {(n-1)!! \over (n+2)!!}\pi  &(n = 2,4,6,...)  \cr
 0 &(n = 1,3,5,...)},
\end{eqnarray} 
we can perform the angular integration in the kernel:
\begin{eqnarray}
  \tilde K(x,y) =   4 G_T \left[ 1 
 - {9 \over 8} \beta G_T(x+y)   
 + {6 \over 5} \beta^2 G_T^2(x^2+y^2+3xy)   \right]
 + {\cal O}(\beta^3).
 \label{kernel1}
\end{eqnarray} 
Then the SD equation reads
\begin{eqnarray}
 B(x) = m_0 + {G_T \over 4\pi^2}
 \int_0^{\Lambda^2}dy {yB(y) \over y+B^2(y)} K(x,y),
\end{eqnarray} 
where we have defined
\begin{eqnarray}
K(x,y):=(4G_T)^{-1}\tilde K(x,y).
\end{eqnarray} 
Since the kernel $K(x,y)$ is symmetric in $x, y$, the
existence of the nontrivial solution is guaranteed by a
mathematical theorem for the integral equation with a
symmetric kernel \cite{Tricomi}, as long as the cutoff
$\Lambda$ is finite.
However, the existence of the solution in the continuum
limit $\Lambda\rightarrow \infty$ is a nontrivial problem
which we address in this paper.
\par
In the limit $\beta \rightarrow 0$, the kernel
$K(x,y)$ is a constant:
$K(x,y) \rightarrow 1$.  Hence the SD equation reduces
to the well-known gap equation whose solution is given by 
a constant (i.e., momentum-independent) mass
$B(p^2) \equiv m$:
\begin{eqnarray}
 m =  m_0 + {G_T \over 4\pi^2} \int_0^{\Lambda^2}dy 
 {ym \over y+m^2}.
\end{eqnarray} 
In the chiral limit $m_0 \rightarrow 0$, this gap equation
is easily solved to give the following scaling law:
\begin{eqnarray}
{1 \over g} := {4\pi^2 \over G_T\Lambda^2} 
 = 1 - {m_d^2 \over \Lambda^2}
\ln \left( 1+ {\Lambda^2 \over m_d^2} \right).
\end{eqnarray} 
This shows that the critical coupling $g_c$ is given by
\begin{eqnarray}
 g_c = 1,
\end{eqnarray} 
and that the critical exponent for the dynamical
fermion mass $m_d$ has the mean-field value,
$1/2$, apart from the logarithmic correction.  To see more
details on the critical behavior, it is convenient to left
$m_0$ behind:
\begin{eqnarray}
 {m_0 \over \Lambda} = g (g^{-1} - 1){m \over \Lambda}
 + g {m^3 \over \Lambda^3}
 \ln \left( 1+ {\Lambda^2 \over m^2} \right).
 \label{eosTh}
\end{eqnarray} 
This is the equation of state for the Thirring model.
The equation of state is able to deduce the critical
exponents, $\nu, \beta, \gamma, \sigma, \delta$ which take
their mean-field values.
\par
Taking account of the terms up to ${\cal O}(\beta)$ in
the kernel, we obtain 
\begin{eqnarray}
 B(x) &=&  m_0 + {G_T \over 4\pi^2}
 \int_0^{\Lambda^2}dy {yB(y) \over y+B^2(y)} \left[ 1 
 - {9 \over 8} \beta G_T y \right]
 \nonumber\\
 && - {G_T \over 4\pi^2}{9 \over 8} \beta G_T x
 \int_0^{\Lambda^2}dy {yB(y) \over y+B^2(y)} .
\end{eqnarray} 
Hence we have
\begin{eqnarray}
 B'(x) =  - {G_T \over 4\pi^2}{9 \over 8} \beta G_T  
 \int_0^{\Lambda^2}dy {yB(y) \over y+B^2(y)} .
\end{eqnarray} 
The RHS is an $x$-independent constant, $-a_1$.
This implies that $B'(x)=-a_1<0$ as long as $B(x)>0$.
As $B''(x) = 0$, the solution is written in the form:
\begin{eqnarray}
 B(p^2) =  - a_1 p^2 + m 
 = m \left( 1-{a_1 \over m}p^2 \right) . 
 \label{gsol4}
\end{eqnarray} 
The solution is determined from the initial conditions:
$B(0)=m$, i.e.,
\begin{eqnarray}
 m =  m_0 +  {G_T \over 4\pi^2}
 \int_0^{\Lambda^2}dy {yB(y) \over y+B^2(y)} \left( 1 
 - {9 \over 8} \beta G_T y \right),
 \label{icb}
\end{eqnarray} 
and $B'(0)=a_1$, i.e.,
\begin{eqnarray}
 a_1 =  {G_T \over 4\pi^2}{9 \over 8} \beta G_T  
 \int_0^{\Lambda^2}dy {yB(y) \over y+B^2(y)} .  
 \label{ica}
\end{eqnarray} 
Therefore, the coefficients $a_1$ and $m$ in
Eq.~(\ref{gsol4}) are determined by solving Eq.~(\ref{icb})
and Eq.~(\ref{ica}).
\par
The critical line in the phase diagram $(\beta, G_T)$ is
obtained from the bifurcation solution
\footnote{
Note that the coefficients $a_1$ and $m$ in the bifurcation
solution are different from those of the solution of the
original nonlinear SD equation, although we use the same
notation. }
$b(p^2) = - a_1 p^2 + m$.
Hence Eq.~(\ref{icb}) and Eq.~(\ref{ica}) are respectively
replaced with
\begin{eqnarray}
 m =   m_0 + {G_T \over 4\pi^2}
 \int_0^{\Lambda^2}dy  b(y)  \left( 1 
 - {9 \over 8} \beta G_T y \right),
 \label{icb'}
\end{eqnarray} 
and 
\begin{eqnarray}
 a_1 =  {G_T \over 4\pi^2}{9 \over 8} \beta G_T  
 \int_0^{\Lambda^2}dy  b(y) .  
 \label{ica'}
\end{eqnarray} 
When $m_0=0$, we can obtain only the ratio of two
coefficients, $a_1/m$ from the bifurcation equation,
Eq.~(\ref{icb'}) and Eq.~(\ref{ica'}).  From the
Eq.~(\ref{ica'}), the ratio is obtained as
\begin{eqnarray}
 {a_1 \over m} \Lambda^2 
 = {{9 \over 8}  \beta_e g^2 \over
 1 + {9 \over 16} \beta_e g^2 },  
\end{eqnarray} 
while, in the chiral limit, Eq.~(\ref{icb'}) leads to
\begin{eqnarray}
 {a_1 \over m} \Lambda^2 
 = 2 {1-g^{-1}- {9 \over 16} \beta_e g
\over 1 - {3 \over 4}  \beta_e g},  
\end{eqnarray} 
where we have introduced the dimensionless coupling
constants:  
\begin{eqnarray}
 \beta_e := {4\pi^2 \over e^2},
 \quad
 g := {G_T\Lambda^2 \over 4\pi^2}.  
\end{eqnarray} 
\par
>From Eq.~(\ref{ica'}), we find that the mass function
behaves as
\begin{eqnarray}
 b(p^2) = m \left[ 1 - {9 \over 8}  \beta_e g^2
 {p^2 \over \Lambda^2} \right] 
 + {\cal O}(\beta_e^2) .
\end{eqnarray} 
Equating two ratios, Eq.~(\ref{icb'}) and
Eq.~(\ref{ica'}), we can obtain the equation of the
critical line. Note that the bifurcation solution gives the
exact critical line.
\par
However this method does not lead to the correct
scaling law in the neighborhood of the critical point of the
Thirring model, i.e., in the limit $\beta \rightarrow 0$.
This is similar to the situation encountered in the
gauged NJL model: the bifurcation solution is
insufficient to derive the correct scaling law of the NJL
model in the $e \rightarrow 0$ limit.   Therefore, we adopt
the Miransky's prescription to linearize the non-linear SD
equation to study the scaling law in the strong
gauge-coupling region
$\beta \ll 1$:
\begin{eqnarray}
 B(x) =   m_0 + {G_T \over 4\pi^2}
 \int_0^{\Lambda^2}dy {yB(y) \over y+B^2(0)} 
 K(x,y).
\end{eqnarray} 
By putting $x=0$, this SD equation leads to
\begin{eqnarray}
 {m_0 \over \Lambda} = {m \over \Lambda} 
 - {g \over \Lambda^3}
 \int_0^{\Lambda^2}dy {yB(y) \over y+m^2} 
 K(0,y).
\end{eqnarray} 
For the kernel~(\ref{kernel1}), it is not difficult to see
that the solution is given in the form:
\begin{eqnarray}
 B(p^2) = m \left[ 1 + \sum_{n=1}^{\infty} (-1)^n a_n
 {(p^2)^n \over m} \right].
\end{eqnarray} 
Once we know all the ratios $a_n/m (n=1,2,...)$ from the
boundary condition, we can write the equation of state.  
\par
Making use of the replacement,
\begin{eqnarray}
  {y \over y+m^2} = 1- {m^2 \over y+m^2},
\end{eqnarray} 
we can write down the equation of state:
\begin{eqnarray}
 {m_0 \over \Lambda} =  g {m \over \Lambda} \tau(\beta,g)
  + g{m^3 \over \Lambda^3} 
  A \left( {m \over \Lambda};\beta,g \right) ,
  \label{eos}
\end{eqnarray} 
where
\begin{eqnarray}
 \tau(\beta_e,g) := g^{-1} - F(\beta_e,g)
 :=  g^{-1} - \int_0^{\Lambda^2}{dy \over \Lambda^2}
 {B(y) \over m} K(0,y),
 \label{deftau}
\end{eqnarray} 
and
\begin{eqnarray}
  A \left( {m \over \Lambda};\beta_e,g \right) 
  :=  \int_0^{\Lambda^2}  {dy \over y+m^2} 
 {B(y) \over m} K(0,y) .
 \label{defA}
\end{eqnarray} 
The general form of $A$ is obtained from
\begin{eqnarray}
  A \left( {m \over \Lambda};\beta,g \right) 
 &=&   \int_0^{\Lambda^2} {dy \over y+m^2} 
 \left[ 1 + \sum_{n=1}^{\infty} (-1)^n 
 {a_n \Lambda^{2n} \over m} 
 \left({y \over \Lambda^2}\right)^n \right]
 \nonumber\\&& \times
 \left[ 1 + \sum_{l=1}^{\infty} (-1)^l K_l
 (\beta G_T y)^l \right].
\end{eqnarray} 
It is not difficult to show that 
$A \left( {m \over \Lambda};\beta,g \right)$
diverges at most logarithmically for large $\Lambda$:
\begin{eqnarray}
A \left( {m_d \over \Lambda};\beta,g \right)
\sim  \ln {\Lambda^2 \over m_d^2} ,
\end{eqnarray} 
since $a_n \Lambda^{2n}/m$ is finite even in the limit 
$\Lambda \rightarrow \infty$.
On the other hand, $F(\beta_e,g)$ is finite  even in the
limit $\Lambda \rightarrow \infty$.
\par
In the chiral limit $m_0 \rightarrow 0$, the equation of
state may have a nontrivial solution, that is to say,
the fermion mass $m_d$ is dynamically generated when
$\tau(\beta_e,g)<0$ (as long as $A>0$). 
Therefore the condition $\tau(\beta_e,g)=0$ gives the
critical line in the two-dimensional phase diagram
$(\beta_e,g)$. Moreover, by the standard argument, this
equation of state is able to show
\cite{BLM90} that the scaling law of the gauged Thirring
model is given by the mean-field type, i.e., almost all the
critical exponents defined in the neighborhood of the
critical point take their mean-field values, apart from the
logarithmic correction coming from $A$.
\par
If we use the ratio obtained up to ${\cal O}(\beta_e)$ in
Eq.~(\ref{ica'}), we get
\begin{eqnarray}
  \tau(\beta_e,g) 
  = g \left[(g^{-1})^2 - g^{-1} + {9 \over 8} \beta_e
\right] 
  + {\cal O}(\beta_e^2),
\end{eqnarray} 
Hence the critical line is obtained as (see Fig.~6)
\begin{eqnarray}
g^{-1} = g_c^{-1}(\beta_e) :=
  1 - {9 \over 8} \beta_e + {\cal O}(\beta_e^2)
 = 1 - {27 \over 8} {\beta_e \over \beta_e^c} 
 + {\cal O}(\beta_e^2),
 \label{criticalline1}
\end{eqnarray} 
where we have used the critical coupling of pure QED,
\begin{eqnarray}
\beta_e^c = 3, \quad i.e. 
\quad {e_c^2 \over 4\pi} = {\pi \over 3} .
\end{eqnarray} 
For example, the
dynamical fermion mass obeys the mean-field type scaling:
\begin{eqnarray}
m_d \sim \Lambda [g_c^{-1}(\beta_e) - g^{-1}]^{1/2}
\end{eqnarray} 
 at fixed $\beta_e \ll 1$. 
These results should be compared with the result of the
effective potential in the previous section.
The result obtained from the SD equation are qualitatively in
good agreement with the previous result.
\par
\par
It is clear from the above treatment how to include more
higher-orders in $\beta_e$.  As all the necessary materials
are also presented above, we do not repeat the same type of
calculation any more.  It is not difficult to show
that the inclusion of higher orders does not change the
qualitative feature of the phase structure.
To make this issue more clear, the detailed numerical
studies will be given in a subsequent paper
\cite{Kondo96}.

%\newpage
\subsection{strong four-fermion coupling:
$M \ll 1$ or $G_T \gg 1$} 

Next, we pay attention to the region of strong
four-fermion coupling, $G_T \gg 1$.
We introduce a new variable:
\begin{eqnarray}
\zeta := k^2 + \mu^2 ,
\quad 
\mu^2 := M^2/\beta
\end{eqnarray} 
to rewrite the function $D_T$ as
\begin{eqnarray}
D_T^{(0)}(k^2)  = {1 \over \beta \zeta}.
\end{eqnarray} 
Then the nonlocal gauge reads
\begin{eqnarray}
 \eta(k^2) &=&   - 2 +
{6 \zeta \over (\zeta-\mu^2)^{3}} \int_{\mu^2}^{\zeta}
d\zeta'  (\zeta'-\mu^2)^2/\zeta' 
\nonumber\\
&=& - 2 +
{3 \zeta \over (\zeta-\mu^2)^{3}} 
\left[ \zeta^2 - 4\mu^2\zeta + 3\mu^4 
+ 2\mu^4\ln {\zeta \over \mu^2} 
\right].
\end{eqnarray} 
Note that $\mu^2/\zeta<1$, when $k^2>0$.  Expanding the
above nonlocal gauge into the power-series in $\mu^2/\zeta$,
we get
\begin{eqnarray}
 \eta(k^2) = 1 - 3 {\mu^2 \over \zeta} 
 + 3 \left( - 3 + 2 \ln {\zeta \over \mu^2} \right)
\left({\mu^2 \over \zeta}\right)^2 
 + {\cal O}\left[\left({\mu^2 \over \zeta}\right)^3\right].
\end{eqnarray} 
This leads to the following expansion:
\begin{eqnarray}
&& D_T^{(0)}(k^2)[4-\eta(k^2)] 
 \nonumber\\
 &=& {3 \over \beta \zeta} + {3M^2 \over \beta^2 \zeta^2}
+ 3 \left( 3 - 2 \ln {\beta \zeta \over M^2} \right)
{M^4 \over \beta^3 \zeta^3}
+ {\cal O}\left({M^6 \over \beta^4 \zeta^4}\right)
 \nonumber\\
 &=& 3 D_T^{(0)}(k^2) + 3 M^2 [D_T^{(0)}(k^2)]^2 
+ {\cal O}\left\{ M^4 [D_T^{(0)}(k^2)]^3 \right\}.
\end{eqnarray} 
Taking into account only the first term in the RHS of
this equation, we are lead to the SD equation:
\begin{eqnarray}
 B(p^2) &=& m_0 + 3 e^2 \int {d^4q \over (2\pi)^4} {B(q^2)
\over q^2 +B^2(q^2)}{1 \over (p-q)^2+\mu^2}.
\label{zerocharge}
\end{eqnarray} 
This is the same SD equation as
that treated in the study of the zero-charge model
\cite{Kondo89}.
\par
First of all, we consider the limit $\mu \rightarrow 0$
with $e^2$ being kept finite and non-zero, which
corresponds to the $g \rightarrow \infty$ limit.   In this
limit the SD equation~(\ref{zerocharge}) reduces to the SD
equation of the quenched planar QED$_4$ in the Landau gauge:
\begin{eqnarray}
 B(p^2) &=& m_0 + 3 e^2 \int {d^4q \over (2\pi)^4} 
 {B(q^2) \over q^2 +B^2(q^2)}{1 \over (p-q)^2} .
\label{SDqQED$_4$}
\end{eqnarray}  
The kernel (\ref{kernel4}) of the SD equation
Eq.~(\ref{SDB3+1}) is given by
\begin{eqnarray}
 \tilde K(x,y) :=  {2 \over \pi}
 \int_0^{\pi} d \vartheta \sin ^{2} \vartheta   
{3e^2 \over (p-q)^2}
= 3e^2 \left[ {\theta(x-y) \over x} + {\theta(y-x) \over y}
\right].
\label{kernelqQED$_4$}
\end{eqnarray} 
Note that the nonlocal gauge reduces to the
Landau gauge in the limit of quenched planar QED$_4$
\cite{KN89}. The solution of the quenched planar QED$_4$ in the
Landau gauge is well-known \cite{Miransky85}.
In the chiral limit $m_0=0$, this equation has a trivial
solution $B(p^2) \equiv 0$ for arbitrary coupling.  
Even in the chiral limit $m_0=0$, however, the SD equation
has a nontrivial solution, if the gauge coupling is allowed
to be strong, i.e., in the region 
$\beta_e := 4\pi^2/e^2 < \beta_e^c$
(supercritical or strong coupling region) where the chiral
symmetry is spontaneously broken and the fermion mass is
dynamically generated. 
\par
The simplest way to obtain the value of the critical
coupling and the explicit form of the nontrivial solution is
to solve the bifurcation equation \cite{Atkinson}
of Eq.~(\ref{SDqQED$_4$}):
\begin{eqnarray}
 b(p^2) &=&  3 e^2 \int {d^4q \over (2\pi)^4} 
 {b(q^2) \over q^2}{1 \over (p-q)^2} .
\label{bifSDqQED$_4$}
\end{eqnarray} 
The nontrivial solution corresponds to the bifurcation
solution of the original nonlinear integral equation.
The critical coupling constant $\beta_e^c$ is obtained
as the bifurcation point of the solution, 
$\beta_e^c  = 3$.
In the neighborhood of the critical point
$\beta_e=\beta_e^c$, the nontrivial bifurcation solution
in the chiral limit $m_0=0$ is given by
\begin{eqnarray}
   b(p^2) 
   = m \left( {m^2 \over p^2}\right)^{1/2}
\sin \left[ {1 \over 2}\rho \ln {p^2 \over m^2} +
\delta_\rho \right]/\sin \delta_\rho,
\label{aqed4}
\end{eqnarray} 
where 
\begin{eqnarray}
 \rho :=  \sqrt{{\beta_e^c \over \beta_e}-1},
\quad
\delta_\rho = \arctan \rho.
\end{eqnarray} 
Here the solution is normalized so that $b(p^2=m^2)=m$.
\par
This solution leads to the following scaling law of
the essential singularity type (Miransky scaling) for the
dynamical fermion mass $m_d$:
\begin{eqnarray}
  {m_d \over \Lambda} =  
 \exp \left[-{n\pi - \delta_\rho \over \rho} \right] ,
  \quad 
  n = 1, 2, ...
  \label{scalingqqed4}
\end{eqnarray} 
The solution with $n=1$ is the ground state and the
solutions with $n \ge 2$ correspond to the excited state
in the sense that the $n=1$ solution gives the lowest
 (CJT) effective potential than other solutions with $n \ge
2$.
\par
Next, we consider the case of non-zero $\mu^2$, but not so
large:  the ratio
\begin{eqnarray}
 r := {\mu^2 \over \Lambda^2} 
 = {e^2 \over G_T\Lambda^2} 
 = {1 \over \beta_e g},
\end{eqnarray} 
is kept finite (and non-zero) in the limit $\Lambda^2
\rightarrow \infty$. Therefore this situation exactly
coincides with that of the zero-charge model treated in
\cite{Kondo89}. The quenched planar QED$_4$ is obtained  as a
limit $\mu \rightarrow 0$.
The results are summarized as follows.
\par
In this case, the kernel (\ref{kernel4}) of the SD equation
Eq.~(\ref{SDB3+1}) is given by
\begin{eqnarray}
 \tilde K(x,y) &:=&  {2 \over \pi}
 \int_0^{\pi} d \vartheta \sin ^{2} \vartheta   
{3e^2 \over (p-q)^2+\mu^2}
\nonumber\\
&=&  {6e^2 \over x+y+\mu^2+\sqrt{(x+y+\mu^2)^2-4xy}}
\nonumber\\
&& \times \left[ 1 + 
{1 \over 3}{\mu^2 \over \sqrt{(x+y+\mu^2)^2-4xy}}
\right].
\label{kernelzero}
\end{eqnarray} 
In ref. \cite{Kondo89,Kondop}, the approximate
analytic solution of the SD equation Eq.~(\ref{zerocharge})
has been obtained  by replacing the kernel with the
simplified one:
\begin{eqnarray}
 \tilde K(x,y) = 3e^2 \left[ 
 {\theta(x-y) \over x+\mu^2} + {\theta(y-x) \over y+\mu^2}
\right].
\label{kernelzero2}
\end{eqnarray} 
In the neighborhood of the critical point, the
(bifurcation) solution in the chiral limit $m_0=0$ is given
by
\begin{eqnarray}
   b(p^2) 
   = m \left( {m^2+\mu^2 \over p^2+\mu^2}\right)^{1/2}
\sin \left[ {1 \over 2}\rho \ln {p^2+\mu^2 \over
m^2+\mu^2} +
\delta_\rho \right]/\sin \delta_\rho,
\label{solzerocharge}
\end{eqnarray} 
The equation of state is obtained as follows.
\begin{eqnarray}
  m_0  = m {1+\rho^2 \over 2\rho}
  \left( {m^2 +\mu^2 \over \Lambda^2 + \mu^2}\right)^{1/2}
\sin \left[ {1 \over 2}\rho 
 \ln {\Lambda^2 +\mu^2 \over m^2 + \mu^2} + \delta_\rho
\right].
\label{eoszerocharge}
\end{eqnarray} 
In the chiral limit $m_0=0$,  we obtain the following
scaling law for the dynamical fermion mass $m_d$:
\begin{eqnarray}
  {m_d \over \Lambda} = \sqrt{  (1 + r) 
 \exp \left[-{2(n\pi - \delta_\rho) \over \rho} \right] 
 - r},
  \quad   n = 1, 2, ...
  \label{scalingzerocharge}
\end{eqnarray} 
It is obvious that, in the limit $\mu \rightarrow 0$, the
solution Eq.~(\ref{solzerocharge}) and the scaling law
Eq.~(\ref{scalingzerocharge}) reduce to those of quenched
planar QED$_4$, i.e., Eq.~(\ref{aqed4}) and
Eq.~(\ref{scalingqqed4}) respectively.
\par
The critical coupling $e^2_c$ (or $\beta_e^c$) is
obtained as the value of parameters at which
$m_d/\Lambda$ vanishes. As a result, the critical coupling
$e^2_c$ (or
$\beta_e^c$) is written as a function of $r$:
$e^2_c=e^2_c(r)$  (or
$\beta_e^c=\beta_e^c(r)$).  
For a given $r$, the critical coupling
$\beta_e^c(r)$ for $r \le 1$ is given by
\begin{eqnarray}
   {\beta_e^c(r) \over \beta_e^c(0)} = (1+\rho^2)^{-1},
   \quad
   \beta_e^c(0) \equiv \beta_e^c ,
   \label{cratio}
\end{eqnarray} 
using a solution $\rho$ of the equation:
\begin{eqnarray}
   {2\rho \over \rho^2-1} 
   = \tan \left[ {\rho \over 2} \ln (1+r^{-1}) \right],
   \label{ceq}
\end{eqnarray}
For a given $r$, if $\rho$ is a solution of
Eq.~(\ref{ceq}), then so is for $-\rho$.
Note that $\rho^2>0$ for a real value of $\rho$. 
Therefore the real solution of Eq.~(\ref{ceq}) leads to
the critical point $\beta_e^c(r)$ such that 
$\beta_e^c(r)/\beta_e^c(0)<1$.
It is shown that the critical value
$\beta_e^c(r)$ is monotonically decreasing in $r$. 
\par
Note that the critical point should locate on the straight
line 
$g^{-1}=r \beta_e$ and this line coincides with the pure QED
axis in the limit $r \rightarrow 0$, and with the pure
Thirring-model axis in the limit $r \rightarrow \infty$.
Therefore a critical point is obtained as an intersection
point of two lines, 
$g^{-1}=r \beta_e$ and $\beta_e=\beta_e^c(r)$ for a given
$r$ (see Table~1).
In Fig.~6, the critical line is drawn
by plotting the intersection point for various $r$.
Incidentally, the numerical solution of the non-linear
SD equation Eq.~(\ref{SDB3+1}) shows
\cite{Kondo89} that
\begin{eqnarray}
   {\beta_e^c(1) \over \beta_e^c(0)} = {0.5 \over 3}
   \sim 0.17 .
\end{eqnarray}

\par
Moreover, it has been shown that the critical scaling law
obeys the mean-field type, namely, the critical exponent
takes the mean-field value except for the possible 
logarithmic correction:
\begin{eqnarray}
m_d \sim \Lambda
(e^2-e^2_c(r))^{\nu} |\log (e^2-e^2_c(r))|^{\nu'}, 
\quad \nu = {1 \over 2}.
\end{eqnarray} 
This implies that 
\begin{eqnarray}
\langle \bar \psi \psi \rangle \sim 
\Lambda^2 m_d \sim
\Lambda^3
(e^2-e^2_c(r))^{\beta_{ch}} 
|\log (e^2-e^2_c(r))|^{\beta_{ch}'}, 
\quad \beta_{ch} = {1 \over 2},
\end{eqnarray} 
since the chiral order parameter is related with the mass
as follows within the approximation
Eq.~(\ref{kernelzero2}):
\begin{eqnarray}
\langle \bar \psi \psi \rangle 
= {1 \over \pi^2}{\beta_e \over \beta_e^c}
(\Lambda^2+\mu^2)[B(\Lambda^2)-m_0].
\end{eqnarray} 
The critical exponent for the logarithmic
correction in strongly coupled gauge theory has not been
known up to now to the author's knowledge, although it is
tried in ref. \cite{Kondop}.
\par
\par
It is rather difficult to study the effect of
higher orders in $M^2$ analytically beyond the lowest order
given in this paper. The numerical investigations will be
given in a subsequent paper \cite{Kondo96}.

\begin{center}
\centerline{Table 1}
 \begin{tabular}{lccc} 
 \hline
  $r$  & $\rho$ & $\beta_e^c(r)/\beta_e^c(0)$ & $g^{-1}$ \\
 \hline
  0    &     0  &   1      &    0     \\
  0.01 &  0.785 &  0.619   &  0.018   \\
  0.05 &  1.020 &  0.491   &  0.074   \\
  0.10 &  1.176 &  0.420   &  0.126   \\
  0.20 &  1.391 &  0.341   &  0.204   \\
  0.30 &  1.557 &  0.292   &  0.263   \\
  0.40 &  1.699 &  0.258   &  0.309   \\
  0.50 &  1.825 &  0.231   &  0.346   \\
  0.60 &  1.941 &  0.210   &  0.378   \\
  0.70 &  2.048 &  0.193   &  0.404   \\
  0.80 &  2.149 &  0.178   &  0.427   \\
  0.90 &  2.244 &  0.166   &  0.447   \\
  1.00 &  2.335 &  0.155   &  0.465   \\
 \hline
 \label{table1}
 \end{tabular}
\end{center}

\newpage
\section{Equi-correlation-length line}
\setcounter{equation}{0}

\subsection{strong gauge-coupling region: $\beta_e \ll 1$}
Fist we discuss the case of $\beta_e \ll 1$.
>From the equation of state, Eq.~(\ref{eos}), we observe
that in the chiral limit $m_0=0$, $ {m_d \over \Lambda}$
satisfies the following equation:
\begin{eqnarray}
 \tau(\beta,g) := g^{-1} - F(\beta, g)
 = -   {m_d^2 \over \Lambda^2} 
  A \left( {m_d \over \Lambda};\beta,g \right) ,
\end{eqnarray} 
In other words, the correlation length $\xi$ defined by
\begin{eqnarray}
  \xi := {\Lambda \over m_d} ,
\end{eqnarray} 
is given as a function of two coupling constants, $\beta$
and $g$.
Therefore the equation 
\begin{eqnarray}
  g^{-1} - F(\beta, g)
 = -   \xi^{-2}
  A \left( \xi^{-1};\beta,g \right) ,
\end{eqnarray} 
gives the line of equi-correlation-length $\xi$ in the bare
parameter space $(\beta, g)$.
The infinite correlation-length limit
$\xi \rightarrow \infty$ corresponds to the critical line,
\begin{eqnarray}
\tau(\beta,g):= g^{-1} - F(\beta, g) = 0,
\end{eqnarray} 
since
$
\xi^{-2} A \left( \xi^{-1};\beta,g \right)
\rightarrow 0
$.  
Up to ${\cal O}(\beta)$, it is obtained by
Eq.~(\ref{criticalline1}).

\subsection{strong four-fermion coupling region}

When $1/g \ll 1$ and $\beta_e$ is in the neighborhood of
$\beta_e$, the equi-$\xi_f:=\Lambda/m_d$ line is obtained
from Eq.~(\ref{scalingzerocharge}) as
\begin{eqnarray}
 g^{-1}  = \beta_e 
 { \exp \left[-{2(n\pi-\delta_\rho) \over \rho} \right] -
 {m_d^2 \over \Lambda^2} \over
 1 - \exp \left[-{2(n\pi-\delta_\rho) \over \rho} \right] },
\label{rcouplingline0}
\end{eqnarray} 
This is drawn in Fig.~7 for various values of $\xi_f$.
In the limit $\xi \rightarrow \infty$, this line gets equal
to the critical line:
\begin{eqnarray}
 g^{-1} = g_c^{-1}(\beta_e)  
 := \beta_e \left\{
 \exp \left[ {2(n\pi-\delta_\rho) \over \rho} \right] 
 - 1 \right\}^{-1}.
\label{criticalline2}
\end{eqnarray} 
Therefore, in the neighborhood of $\beta_e=\beta_e^c$, the
critical line behaves as
\begin{eqnarray}
 g_c^{-1}(\beta_e)  
 \cong \beta_e  
 \exp \left[-{2(n\pi-\delta_\rho) \over \rho} \right] .
\end{eqnarray} 
All the derivatives of the critical line at
$\beta_e=\beta_e^c$ are zero, since the critical point
$\beta_e=\beta_e^c$ is the essential singularity.
\par
Here it should be noted that the gauged Thirring model has
the dynamically generated fermion mass even if $1/g<0$.
This is interpreted as follows.
Although the original four-fermion coupling is negative
for $1/g<0$, namely, the four-fermion interaction is
repulsive,  a pair of fermion and antifermion can be
condensed to make the bound state due to strong gauge
coupling $\beta_e < 1$. 
It is known that such a phenomenon occurs also in the gauged
NJL model: the dynamical fermion mass generation is possible
even for negative four-fermion coupling, if the gauge
coupling is sufficiently strong \cite{Kondo91a,Kondo91b}.
\par
Therefore this result suggests that the dynamical fermion
mass generation is possible in the negative
four-fermion coupling region
\begin{eqnarray}
G_T < G_T^{c'}(\beta) < 0,
\end{eqnarray} 
as well as the positive coupling region: 
\begin{eqnarray}
G_T > G_T^c(\beta) > 0.
\end{eqnarray} 
It is worth recalling the fact \cite{Klaiber68,IKSY95}
that the 1+1 dimensional Thirring model indeed exhibits the
dynamical mass generation when $G_T > 0$ and $G_T < -\pi$,
i.e.,
$G_T^c(0)=0$ and $G_T^{c'}(0)=-\pi$.
In the gauged Thirring model, the critical values $G_T^c$,
$G_T^{c'}$ depend on the gauge coupling $\beta$.
This issue will deserve further studies.

\subsection{the whole critical line}
We have shown the existence of the critical lines which
extend into the interior of the phase diagram $(\beta,G_T)$
from two critical points:
$(0,G_T^c)$ and $(\beta_c,0)$.
In the neighborhood of the respective critical
point, it was shown that the scaling behavior is consistent
with the mean-field theory, except for possible logarithmic
corrections.
We expect that the two critical lines meet together and
constitutes one critical line as a whole.  Then the critical
line  connects two critical points of pure Thirring model
and pure QED$_4$. 
\par
In order to show this, we need to solve the SD equation
Eq.~(\ref{SDB3+1}) numerically in the nonlocal
gauge~(\ref{nlg3+1}) without any further approximation.
For this, it is necessary to perform
double integration over the angular variable $\vartheta$ and
the radial variable $p^2$. 
Such kind of numerical calculations were done in strong
coupling QED$_4$ beyond the quenched approximation where the
angular integration is a nontrivial problem, see
\cite{KN92,KMN92}. The result of numerical calculation will
be given in a subsequent paper together with other things.

%\newpage
\section{RG flow and the continuum limit}
\setcounter{equation}{0}

\subsection{renormalized coupling constant}

By virtue of the reformulation of the Thirring model as a
gauge theory, the original four-fermion contact interaction
has been replaced with the gauge interaction mediated by
exchanging the gauge boson, see Fig.~8. 
Therefore, in analogy with QED, we can define the {\it
renormalized coupling constant} $\alpha_R$ between the
fermion and the gauge boson  in the chiral symmetry breaking
phase as follows.
\footnote{
In the chiral symmetry breaking phase, the fermion becomes
massive, i.e. the fermion has the pole mass $m_d$ at
$p^2=m_d^2$. Therefore, we can define the renormalized
coupling constant at another point: $k^2=2m_d^2$
for $k^2:=(p-q)^2=p^2+q^2 -2 pq \cos \vartheta$.
However, this changes of the definition does not
essentially influence the result, see the next subsection. 
}
\begin{eqnarray}
 \alpha_R 
 := {1 \over 4\pi^2} [k^2 D_T(k)] |_{k^2=m_d^2}
= {1 \over 4\pi^2} {k^2 \over \beta k^2 + M^2}
\Big|_{k^2=m_d^2} 
= {g \over \beta_e g + {\Lambda^2 \over m_d^2}},
\label{rcoupling}
\end{eqnarray} 
where the factor $1/4\pi^2$ has been introduced to simplify
the final expression.   
Note that this definition of the renormalized
coupling constant does not depend on the
longitudinal part of the gauge boson propagator and hence
is gauge-independent.  
\footnote{
We identify the dynamical fermion mass $m_d$ as a pole
mass which should be gauge-independent, although we
actually use $m_d$ obtained from the approximate SD
equation in the nonlocal gauge.
}
The above equation is rewritten as
\begin{eqnarray}
 {m_d^2 \over \Lambda^2} 
 = {\alpha_R \over g(1- \alpha_R \beta_e)}.
\label{rcouplingm}
\end{eqnarray} 

\subsubsection{strong gauge-coupling limit $\beta_e
\rightarrow 0$}

In the limit $\beta_e \rightarrow 0$,
Eq.~(\ref{rcoupling}) and Eq.~(\ref{rcouplingm}) reduce to
\begin{eqnarray}
 \alpha_R  =  g  {m_d^2 \over \Lambda^2},
 \quad {\rm and} \quad
 {m_d \over \Lambda} = (\alpha_R g^{-1})^{1/2}.
\label{rcoupling'}
\end{eqnarray} 
Substituting this relation into the equation of state
Eq.~(\ref{eosTh}) for the Thirring model, we obtain the
equation of equi-$\alpha_R$ line (in the presence of the
bare fermion mass):
\begin{eqnarray}
 {m_0 \over \Lambda} = g (\alpha_R g^{-1})^{1/2}
 \left[ g^{-1} - 1
 +  \alpha_R g^{-1}
 \ln \left( 1 + \alpha_R g^{-1} \right) \right].
\end{eqnarray} 
In the limit $m_0=0$, the $\alpha_R$ is implicitly given as
a function of $g$ and vice versa:
\begin{eqnarray}
  g^{-1} - 1 +  \alpha_R g^{-1}
 \ln \left( 1 + \alpha_R g^{-1} \right) = 0.
\end{eqnarray} 
If a value of $\alpha_R$
is specified, the corresponding value of $g$ and hence
$m_d/\Lambda$ is uniquely determined according to the
equation of state which is derived as a consequence of SD
equation.   In Table~2, the values of
$g$ and
$m_d/\Lambda$ are enumerated for various $\alpha_R$.  
In the phase diagram, an equi-$\alpha_R$ line on which
$\alpha_R$ has a specified value of $\alpha_R$ begins at the
point on the line
$\beta_e=0$ (pure Thirring-model axis), and smoothly meets
with the piece of the corresponding line in the region
$\beta_e \sim \beta_e^c$, which is obtained in the next
subsection.

\begin{center}
\centerline{Table 2}
 \begin{tabular}{lccc} 
 \hline
  $\alpha_R$  & $g$ & $g^{-1}$ & $m_d/\Lambda$ \\
 \hline
  0    &     1  &   1      &    0     \\
  0.1 &  1.0094 &  0.99064   &  0.31475   \\
  0.2 &  1.0353 &  0.96588   &  0.43952   \\
  0.3 &  1.0739 &  0.93118   &  0.52854   \\
  0.4 &  1.1219 &  0.89129   &  0.59709   \\
  0.6 &  1.2372 &  0.80826   &  0.69639   \\
  1.0 &  1.5085 &  0.66289   &  0.81418   \\
  2.0 &  2.2655 &  0.44140   &  0.93957   \\
  3.0 &  3.0532 &  0.32753   &  0.99125   \\
 \hline
 \label{table2}
 \end{tabular}
\end{center}

\subsubsection{strong four-fermion coupling near
$\beta_e^c$}

When $\beta_e$ is near $\beta_e^c$ 
and $1/g \ll 1$,
the dynamical fermion mass $m_d$ obeys the
following scaling law Eq.~(\ref{scalingzerocharge}):
\begin{eqnarray}
 {m_d^2 \over \Lambda^2} 
 =  (\beta_e g)^{-1} \left\{ (\beta_e g+1) 
 \exp \left[- {2(n\pi-\delta_\rho) \over \rho} \right] -1 
\right\}.
\label{mscaling}
\end{eqnarray} 
Therefore, substituting Eq.~(\ref{mscaling}) into
Eq.~(\ref{rcouplingm}) or Eq.~(\ref{rcoupling}), we get
the equation for the equi-$\alpha_R$ line in the region
$1/g \ll 1$:
\begin{eqnarray}
 g^{-1}
 =  \beta_e \left\{ (1- \alpha_R \beta_e)^{-1} 
 \exp \left[  {2(n\pi-\delta_\rho) \over \rho} \right] -1 
\right\}^{-1}.
\label{rcouplingline}
\end{eqnarray} 
In Fig.~9, this line is drawn for various values of
$\alpha_R$.  The graph shows that the line passes
through the negative $1/g$ region for relatively large value
of
$\alpha_R$ ($\alpha_R > 0.3$).
Especially, in the limit $\alpha_R \rightarrow 0$, the
line of equi-$\alpha_R$ Eq.~(\ref{rcouplingline}) reduces
to Eq.~(\ref{criticalline2}).
As Eq.~(\ref{criticalline2}) is nothing but the critical
line,  the renormalized coupling constant vanishes on the
critical line $m_d/\Lambda=0$. Hence this result seems to
show that the continuum limit of the cutoff gauged Thirring
model is trivial. However there is an exceptional point,
i.e., 
$(\beta_e, 1/g)=(\beta_e^c, 0)$.  All the other
renormalization group flows with non-zero $\alpha_R$ 
converge to this point.  This shows that we can adjust the
bare parameters so that the cutoff gauged Thirring model
moves along the RG flow with $\alpha_R\not=0$ in the
continuum limit and the resulting continuum theory obtained
in such a way has the nontrivial interaction.

\subsection{constant mass ratio}

In order to get further insight into the renormalized
properties of the gauged Thirring model, we study the line
of equi-mass-ratio between the fermion mass and
the gauge-boson mass.
As the mass of the gauge boson is given by
\begin{eqnarray}
 m_A^2 = e^2 G_T^{-1},
\end{eqnarray} 
within our approximation, the mass
ratio reads
\begin{eqnarray}
 r_m := {m_\psi^2 \over m_A^2}
 = {m_d^2 \over e^2 G_T^{-1}}
 = \beta_e g {m_d^2 \over \Lambda^2} .
\label{massratio}
\end{eqnarray} 
As the physical mass is gauge-independent, this ratio
should be also gauge independent.  In our treatment, this
is approximately satisfied, since we use the
fermion mass obtained from the approximate solution of
the SD equation.

\par
If we use
Eq.~(\ref{mscaling}) for the dynamical fermion mass, we get
the line of equi-$r_m$ (mass ratio), see Fig.~10:
\begin{eqnarray}
 g^{-1}
 =  \beta_e \left\{ (1 + r_m)
 \exp \left[ {2(n\pi-\delta_\rho) \over \rho} \right] -1 
\right\}^{-1}.
\label{cmassline}
\end{eqnarray} 
The line (\ref{cmassline}) with $r_m=0$ is equal to the
critical line. We will be able to obtain the continuum
gauged Thirring theory with massive fermion and massive
gauge boson by approaching the point 
$(\beta_e, 1/g)=(\beta_e^c, 0)$ along this line with
$r_m \not= 0$. 
\par
Comparing Eq.~(\ref{rcouplingline}) with
Eq.~(\ref{cmassline}), we can see that the equi-$r_m$
line coincides with the equi-$\alpha_R$ line, if they are
reparameterized according to the following relationship:
\begin{eqnarray}
 r_m
 =  {\beta_e \alpha_R \over 1 -  \beta_e \alpha_R}.
\label{relation}
\end{eqnarray} 
In the limit of approaching the critical point, the above
relationship reduces to
\begin{eqnarray}
 r_m^c
 =  {\beta_e^c \alpha_R \over 1 -  \beta_e^c \alpha_R},
 \quad {\rm or} \quad
 \alpha_R = (\beta_e^c)^{-1} {r_m^c \over r_m^c+1}.
\label{relationc}
\end{eqnarray} 
Here note that $r_m^c$ is monotonically increasing in
$\alpha_R$. Therefore, for the mass ratio to be positive,
$0 < r_m^c \le +\infty$, in the continuum limit, the
$\alpha_R$ should lie in the region:
\begin{eqnarray}
0 < \alpha_R \le (\beta_e^c)^{-1}  = {1 \over 3}.
\label{region}
\end{eqnarray} 
Therefore, the renormalized coupling constant $\alpha_R$
should have an upper bound, for the continuum theory to
be physically meaningful. 
\footnote{
If we adopt an alternative definition of the renormalized
coupling constant,  the relation (\ref{relation}) changes as
follows.
$$
 r_m
 =   {1 \over 2}
 {\beta_e^c \alpha_R \over 1 -  \beta_e^c \alpha_R},
 \quad {\rm or} \quad
 \alpha_R = (\beta_e^c)^{-1} {r_m \over r_m+1/2}.
$$
However, this does not change any conclusion, e.g.
Eq.~(\ref{region}).
}
In this range of $\alpha_R$, the equi-$\alpha_R$ lines
locate in the positive region, $1/g>0$.

\subsection{scaling of the mass function}

Note that the bifurcation solution (\ref{solzerocharge}) is
rewritten as
\begin{eqnarray}
   {b(p^2) \over m_d}
=  \left( 
{1+r_m^{-1} \over {p^2 \over m_d^2}+r_m^{-1}}\right)^{1/2}
\sin \left[ {1 \over 2}\rho \ln 
{{p^2 \over m_d^2}+r_m^{-1} \over 1+r_m^{-1}} +
\delta_\rho \right]/\sin \delta_\rho ,
\end{eqnarray} 
where we have used
\begin{eqnarray}
   {\mu^2 \over m_d^2} 
   = {\Lambda^2 \over m_d^2}{1 \over \beta_e g}
   = r_m^{-1}.
\end{eqnarray} 
This shows that the fermion mass function has the universal
form on the equi-$r_m$ line and hence on the equi-$\alpha_R$
line.  Therefore, if we plot the fermion mass function
$b(p^2)/m_d$ as a function of $p/m_d$, the curves agree
for various different values of the cutoff (or
$m_d/\Lambda$) on the equi-$\alpha_R$ line.  Hence the
scaling holds in the neighborhood of the critical point
$(\beta_e,1/g)=(\beta_e^c,0)$.

\subsection{beyond the quenched approximation}

In this paper we have adopted the quenched approximation
to obtain the analytical solution of the SD equation for
the fermion mass function.  And, we have
inevitably substituted the result under our
approximations into the
$m_d/\Lambda$ of Eq.~(\ref{rcoupling}) and
(\ref{massratio}) to obtain the RG flows.  However, the
following result is unchanged even if we use the exact
solution of the SD equation to study the RG flows:  it is
necessary that the four-fermion coupling constant
approaches infinity $g(\Lambda) \rightarrow \pm \infty$ for
obtaining the nontrivial continuum theory in the continuum
limit $\Lambda
\rightarrow \infty$. In fact, if $g$ remains
finite in the continuum limit $\Lambda \rightarrow \infty$, 
both $\alpha_R$ and $r_m$ go to zero in this limit.
Therefore, if we want to obtain the nonzero  $\alpha_R$ and
$r_m$, $g$ must go to infinity while $\beta_e$
must converge some finite value.
Therefore, although the explicit RG flows may change
depending on the approximation, the above scenario for
obtaining the nontrivial continuum limit should hold
irrespective of the approximation.

%\newpage
\section{Anomalous dimensions}
\setcounter{equation}{0}

\subsection{strong gauge-coupling region}
First, we obtain the wavefunction renormalization
constant defined by
\begin{eqnarray}
 Z_m := {\partial m_0 \over \partial m} \Big|_{m=m_d} .
\end{eqnarray} 
>From the equation of state, Eq.~(\ref{eos}), we observe
that in the chiral limit $m_0=0$
\begin{eqnarray}
 \tau(\beta,g) := g^{-1} - F(\beta, g)
 = -   {m_d^2 \over \Lambda^2} 
  A \left( {m_d \over \Lambda};\beta,g \right) .
\end{eqnarray} 
>From the equation of state Eq.~(\ref{eos}), we have
\begin{eqnarray}
 {\partial m_0 \over \partial m} 
 = g \tau(\beta,g)
  + 3 g{m^2 \over \Lambda^2} 
  A \left( {m \over \Lambda};\beta,g \right) 
  + g {m^3 \over \Lambda^3} 
  A' \left( {m \over \Lambda};\beta,g \right)  ,
\end{eqnarray} 
where the prime in the third term of RHS implies the
differentiation with respect to $m/\Lambda$.
The third term is calculated as follows.
\begin{eqnarray}
 A' \left( {m \over \Lambda};\beta,g \right)
&:=& {\partial \over \partial m/\Lambda} 
 A \left( {m \over \Lambda};\beta,g \right)
 \nonumber\\
 &=&  - 2{\Lambda^3 \over m}
 {\partial \over \partial \Lambda^2} 
 A \left( {m \over \Lambda};\beta,g \right)
 \nonumber\\
&=&  - 2{\Lambda^3 \over m}
{1 \over \Lambda^2+m^2} {B(\Lambda^2) \over m}
K(0,\Lambda^2) ,
\end{eqnarray} 
where we have used Eq.~(\ref{defA}).
Therefore we obtain
\begin{eqnarray}
 Z_m &=& - 2 g \tau(\beta,g) - 2 g {m_d^2 \over \Lambda^2}
\left(1+{m_d \over \Lambda^2} \right)^{-1}
{B(\Lambda^2) \over m_d}K(0,\Lambda^2) 
\nonumber\\
 &=&   2 g \left[ {m_d^2 \over \Lambda^2}
A \left( {m_d \over \Lambda};\beta,g \right) 
-   {m_d^2 \over \Lambda^2}
\left(1+{m_d \over \Lambda^2} \right)^{-1}
{B(\Lambda^2) \over m_d}K(0,\Lambda^2) \right]
\nonumber\\
 &=&  2 {m_d^2 \over \Lambda^2}
\left[ F(\beta,g) - {m_d^2 \over \Lambda^2}
A \left( {m_d \over \Lambda};\beta,g \right) \right]^{-1}
\nonumber\\&& \times
\left[ A \left( {m_d \over \Lambda};\beta,g \right)
- \left(1+{m_d \over \Lambda^2} \right)^{-1}
{B(\Lambda^2) \over m_d} K(0,\Lambda^2) \right] .
\label{wfrc}
\end{eqnarray} 
In the limit $\beta \rightarrow 0$, especially, we obtain
for large $\Lambda/m$
\begin{eqnarray}
 Z_m =   2 {m_d^2 \over \Lambda^2}
\left[ 1 - {m_d^2 \over \Lambda^2}
\ln {\Lambda^2 \over m_d^2} \right]^{-1}
\left[ \ln {\Lambda^2 \over m_d^2} - 1 \right].
\end{eqnarray} 

\par
Now we calculate the anomalous dimension of the
composite operator $\bar \psi \psi$ according to the formula
\cite{CG89}:
\begin{eqnarray}
 \gamma_{\bar \psi \psi} := 
 - {\partial \ln Z_m \over \partial \ln \Lambda}
\Big|_{\Lambda/m \rightarrow \infty} .
\label{adimdef}
\end{eqnarray} 
>From Eq.~(\ref{wfrc}), we obtain for large $\Lambda/m$
\begin{eqnarray}
&&  - {\partial \ln Z_m \over \partial \ln \Lambda} 
\nonumber\\ 
 &=&   2 +  \left[ F - {m_d^2 \over \Lambda^2}
A \left( {m_d \over \Lambda} \right) \right]^{-1}
{\partial \over \partial \ln \Lambda}
\left[ F - {m_d^2 \over \Lambda^2}
A \left( {m_d \over \Lambda} \right) \right]
\nonumber\\&&
- \left[ A \left( {m_d \over \Lambda} \right)
- {B(\Lambda^2) \over m_d} K(0,\Lambda^2) \right]^{-1} 
{\partial \over \partial \ln \Lambda}
\left[ A \left( {m_d \over \Lambda} \right) -
{B(\Lambda^2) \over m_d} K(0,\Lambda^2) \right]  ,
\end{eqnarray} 
where we have omitted to write the dependence of the
coupling constants.
Note that $F(\beta,g)$ and
${B(\Lambda^2) \over m_d} K(0,\Lambda^2)$ are 
$\Lambda$-independent constants, as discussed in section
9.  Hence, the derivative 
\begin{eqnarray}
{\partial  \over \partial \ln\Lambda}
A \left( {m_d \over \Lambda};\beta,g \right)
= 2 {\Lambda^2 \over \Lambda^2+m^2} 
{B(\Lambda^2) \over m_d} K(0,\Lambda^2)
\end{eqnarray} 
is finite in the limit $\Lambda \rightarrow \infty$.
This is equivalent to say that $A$ diverges at most
logarithmically as the cutoff $\Lambda$ increases.
Therefore the anomalous dimension $\gamma_m$ defined by
Eq.~(\ref{adimdef}) approaches the large value in the
continuum limit $\Lambda \rightarrow \infty$:
\begin{eqnarray}
 \gamma_{\bar \psi \psi} \rightarrow  2 ,
\label{adim}
\end{eqnarray} 
irrespective of the values of the coupling constants,
$\beta, g$.

%\newpage
\subsection{strong four-fermion coupling region}
>From the equation of state Eq.~(\ref{eoszerocharge}), 
we obtain the wavefunction renormalization:
\begin{eqnarray}
 Z_m  =  {1+\rho^2 \over 2}
 {m_d^2 \over m_d^2 + \mu^2}
\left( {m_d^2 +\mu^2 \over \Lambda^2 + \mu^2}\right)^{1/2} 
\cos \left[ {1 \over 2}\rho 
 \ln {\Lambda^2 +\mu^2 \over m_d^2 + \mu^2} + \delta_\rho
\right].
\end{eqnarray} 
Then we have
\begin{eqnarray}
&&  - {\partial \ln Z_m \over \partial \ln \Lambda} 
\nonumber\\ 
 &=&   2 -  {m_d^2 \over m_d^2+\mu^2}
 + \rho  {m_d^2 \over m_d^2+\mu^2}
\tan \left[ {1 \over 2}\rho 
 \ln {\Lambda^2 +\mu^2 \over m_d^2 + \mu^2} + \delta_\rho
\right].
\end{eqnarray} 
If we substitute the scaling law
Eq.~(\ref{scalingzerocharge}) for the dynamical fermion mass
$m_d$ into the RHS of this equation, the third term vanishes, since the argument of
$\tan$ is $n\pi (n=1,2,...)$.
Hence the anomalous dimension reads
\begin{eqnarray}
  \gamma_{\bar \psi \psi} 
=  2 -  {m_d^2 \over m_d^2+\mu^2}
= 2 - {\beta_e g m_d^2/\Lambda^2 \over 
1 + \beta_e g m_d^2/\Lambda^2}.
\label{anodim}
\end{eqnarray} 
By putting $\mu=0$, Eq.~(\ref{anodim}) correctly reproduces
the large anomalous dimension of quenched QED$_4$
\cite{Miransky85,MY89}:
\begin{eqnarray}
\gamma_{\bar \psi \psi} = 1.
\end{eqnarray} 
If we move the cutoff gauged Thirring model to
the critical point in the arbitrary way, i.e.
irrespective of the RG flows, we get the largest anomalous
dimension,
$\gamma_m=2$, as in the case of the previous section.
However this is physically meaningless, since this leads
to the trivial continuum limit, i.e. $\alpha_R=0$.
\par
If the continuum limit $\Lambda \rightarrow
\infty$ of the gauged Thirring model is taken along the RG
trajectories obtained in the previous section, the anomalous
dimension is related to both the renormalized coupling
constant and the mass ratio as follows.
\begin{eqnarray}
  \gamma_{\bar \psi \psi} 
= {2+r_m \over 1+r_m} = 2 - 3 \alpha_R .
\label{anorel}
\end{eqnarray} 
This relation implies that the gauged Thirring model has 
the anomalous dimension definitely between 1 and 2 in the
continuum limit:
\begin{eqnarray}
  1< \gamma_{\bar \psi \psi} < 2.
\end{eqnarray} 
In other words, the anomalous dimension of the continuum
limit of the gauged Thirring model is uniquely determined by
specifying the RG trajectory along which the continuum
limit is taken. The equi-$\gamma_m$ line is obtained by
reparametrizing the equi-$\alpha_R$ line or equi-$r_m$ line
according to Eq.~(\ref{anorel}).
The large anomalous dimension $\gamma_{\bar \psi \psi}>1$
implies the physical dimension of the four-fermion
operator gets less than four in 3+1 spacetime dimensions:
$
{\rm dim}[(\bar \psi \psi)^2]
= 2(3-\gamma_{\bar \psi \psi}) < 4
$.
This implies that the four-fermion interaction becomes
relevant, which will lead to the (nonperturbative)
renormalizability of the four-fermion interaction.

\newpage
\section{Conclusion and discussion}
\setcounter{equation}{0}

In this paper we have proposed a gauge-invariant
generalization of the Thirring model, i.e. four-fermion
interaction of the current-current type.  We call this
model the gauged Thirring model.
The original Thirring model is recovered as the strong
gauge-coupling limit of the gauged Thirring model. 
This is in sharp contrast with the gauged NJL model in which
the NJL model is obtained as the weak gauge-coupling limit.
\par
In the chiral limit $m_0=0$, it has been shown that there
occurs the spontaneous breaking of the chiral symmetry.  We
have obtained the critical line which separates the
chiral-symmetry-breaking phase from the chiral-symmetric
phase.  Moreover, we have shown that the phase transition
associated with the spontaneous chiral-symmetry breaking is
of the second order by examining the gauge-invariant
effective potential for the chiral order parameter.
In order to study the phase structure in more detail, we
have written down the SD equation for the fermion
propagator in the nonlocal $R_\xi$ gauge and obtained the
bifurcation solution of the SD equation in the two regions
which are neighborhoods of the critical points of the pure
Thirring model and the  pure QED$_4$.  Both approaches have
given the same values for critical exponents and the
same result on the order of the chiral transition. 
\par
It is worth recalling that the introduction of the gauge
interaction to the NJL model of four-fermion interaction
has promoted the trivial and perturbatively
non-renormalizable NJL model into the nontrivial
interacting model in the continuum limit
\cite{KSY91,Yamawaki92,KTY93,KSTY94,HKKN94}.   
We have searched for such a possibility also in the gauged
Thirring model. Actually we have obtained the RG flows
(lines of constant physics) for the gauged Thirring model.
Proposing the gauge-invariant criterion of triviality or
nontriviality, we have examined the continuum limit of the
cutoff gauged Thirring model and obtained a signal of the
nontrivial continuum limit.   The nontrivial continuum
gauged Thirring theory with large anomalous dimension
between 1 ant 2 will be obtained by approaching the critical
point
$(\beta,g)=(\beta^c,\pm \infty)$ along the RG trajectories
of constant renormalized coupling. 
This way of approaching the critical point determines
uniquely the ratio of the fermion mass to the gauge boson
mass in the continuum limit.
These results are based on the approximate
solution of the SD equation for the fermion propagator. 
It will be the next subject to study whether the
nontriviality of the gauged Thirring model survives after
the improvement of the approximation. 
\par
There is a possibility of understanding the nontriviality
of the gauged Thirring theory from another point of view.
This is based on the observation that the gauged Thirring
model may be obtained as a special limit of the
perturbatively-renormalizable extended model of the gauged
Thirring model, in the similar way that the gauged NJL
model is obtained as a limit of gauge-Higgs-Yukawa
system \cite{HKKN94}.  
For example, such an extended model will be
given by
\begin{eqnarray}
 {\cal L}_{gTh}  
 &=&  \bar \psi (i \gamma^\mu \partial_\mu - m) \psi
 + A_\mu \bar \psi \gamma^\mu \psi
  \nonumber\\
&& + {\beta \over 4} F^{\mu\nu}F_{\mu\nu} 
+ {1 \over 2G_T}  |(\partial_\mu - i A_\mu) \phi |^2
 + m_\phi^2 |\phi |^2 + \lambda (|\phi |^2)^2 .
 \label{ggTh}
\end{eqnarray}
The behavior of this model is well known in the weak
coupling region:
$e \ll 1 (\beta \gg 1)$ 
in which the perturbation theory in the coupling
constant $e^2$ is applicable well, see
\cite{CW73,Yamagishi81}. However, one does not know much of
the behavior of this model in the strong coupling region:
$e \gg 1 (\beta \ll 1)$.
For the gauged Thirring model, therefore, the usual
RG function obtained up to some finite order of perturbation
theory is not necessarily powerful in sharp contrast with
the gauged NJL model, since we are dealing with the
very strong gauge-coupling region. Therefore, we need some
non-perturbative methods.
\par
The lattice theory is one of the possibilities.
In relation with this, we must mention the recent work of
Jersak et al. \cite{FJ95,FFJL95} in which a
gauge-invariant generalization of the NJL model was
proposed.  In our opinion, their model is nothing but the
lattice version of the gauged Thirring model.  They
start from the lattice formulation with the compact gauge
field which will be necessary to extend the model into the
non-Abelian case. However, their result shows very
complicated phase structures which make the analysis of
the continuum limit quite difficult. 
This may be an artifact coming from the compact
formulation of the Abelian gauge theory and hence the
comparison with the non-compact formulation will be needed
on the lattice.   However, their studies are quite
interesting, if their model is identified with a simplified
version of the  non-Abelian gauged Thirring model.
\par
Actually the extension of the gauged Thirring model to the
non-Abelian case is quite important, since such a model
will be a low energy effective theory of QCD.  Our
formulation of the gauged Thirring model is superior to
the previous formulation
\cite{ER86,AR95} in a point that our formulation respects
the gauge invariance  from the beginning as well as all the
symmetries of the original QCD. The analysis of the
non-Abelian case will be given in one of the subsequent
works \cite{Kondo96}.
\par
Another purpose of this work is to clarify
the gauge-invariant mechanism for mass
generation so that it can be replaced with the usual
Higgs mechanism. In the very massive case $m_0 \gg 1$, we
can shed light on this problem through the procedure of
bosonization. 
Actually, the 2+1 dimensional massive Thirring
model can be bosonized into the Maxwell-Chern-Simons theory
in which the gauge boson mass is generated through the
Chern-Simons term without breaking the gauge invariance
\cite{FS94,Kondo95Th2}. This scenario can be extended into
the higher dimensional case
\cite{Kondo95Th2,IKN95}.   
In 3+1 dimensional case, it is shown \cite{IKN95} that the
Thirring model can be mapped into the antisymmetric tensor
gauge theory.  And the mass for the tensor gauge field is
given in the gauge-independent way.
However, we need further studies to clarify the
non-perturbative renormalizability of the tensor gauge
theory preserving the unitary. 
This subject will be discussed elsewhere.
\par
Finally, we mention other possibilities of order
parameters.  It is very interesting to study the 
gauge-invariant (neutral) fermion field
$F(x) := \phi^\dagger(x) \psi(x) $ and
$\bar F(x) := \bar \psi(x) \phi(x)$.
Although the neutral fermion gives the same local order
parameter as the charged fermion $\psi$:
\begin{eqnarray}
 \langle \bar F(x) F(x) \rangle
 = \langle \bar \psi(x) \psi(x) \rangle ,
\end{eqnarray}
they may have the different two-point functions:
\begin{eqnarray}
 \langle \bar F(x) F(y) \rangle
 = \langle \bar \psi(x) \phi(x) \phi^\dagger(y) \psi(y)
\rangle .
\end{eqnarray}
This suggests that the neutral fermion mass $m_F$ may be
used as a different nonlocal order parameter.  If we
regard that
$m_F = 0$ implies $m_\psi = 0$, we can see that
$m_\psi \not= 0$ implies $m_F \not= 0$. 
However, $m_\psi = 0$
leads to two possibilities:  
$m_F = 0$, or $m_F \not= 0$.  
Hence the chiral-symmetric phase  with
$\langle \bar \psi(x) \psi(x) \rangle=0$
can be distinguished by the neutral fermion mass $m_F$, as
suggested by Jersak et al. \cite{FJ95}.
Moreover, it is also interesting to make clear
the interweaving of the chiral and Higgs phase transition
\cite{FJ95}.
There are a lot of questions which are waiting to be
clarified in the future works \cite{Kondo96}.

\section*{Acknowledgments}
The author would like to thank Gerrit Schierholz and all
members of HLRZ for warm hospitality in HLRZ,
Forschungzentrum KFA J\"ulich as a research fellow of the
Alexander von Humboldt Foundation. He is also very grateful
to Jiri Jersak for valuable discussions on the dynamical
fermion mass generation by strong gauge interaction
shielded by a scalar field. This work is supported in part 
by the Alexander von Humboldt Foundation and
the Grant-in-Aid for Scientific Research from the Ministry
of Education, Science and Culture (No.07640377).

\newpage
\appendix
\section{Inversion method}
\setcounter{equation}{0}

\subsection{general strategy}
\par
First we add the source term $S_J$ into the original
action 
$S_0 :=\int d^D x {\cal L}_0[\phi_i(x)]$
where $\phi_i$ denotes various fields collectively.  
For example, in order
to calculate the local order parameter 
$\varphi_J(x) := \langle O(x) \rangle$,
we choose the source term:
\begin{eqnarray}
S_J := \int d^D x {\cal L}_J = \int d^D x J(x) O(x),
\end{eqnarray}
where $J$ is the artificial source.
Then we calculate the vacuum action (generating
functional of the connected Green's functions):
\begin{eqnarray}
 W[J] = - \ln Z[J] .
\end{eqnarray}
where $Z[J]$ is the partition function
in the presence of the source:
\begin{eqnarray}
 Z[J] :=  \int [d\phi_i(x)] \exp \left[ 
 -(S_0[\phi_i]+S_J[O]) \right].
\end{eqnarray}
 The order parameter 
$\varphi_J(x)$ is calculated for the
theory with the action:
$S=S_0+S_J$:
\begin{eqnarray}
 \varphi_J(x) \equiv {\delta W[J] \over \delta J(x)} .
\end{eqnarray}
It should be noted that the order parameter $\varphi_J$ (as
a function of $J$) is non-zero for
$J\not=0$ and vanishes in the limit $J \rightarrow 0$ at
this stage.
\par
The effective action $\Gamma[\varphi]$
is defined through the Legendre transformation:
\begin{eqnarray}
  \Gamma[\varphi] 
  := W[J] - \int d^D x J(x)  \varphi_J(x) .
\end{eqnarray}
Thus we have the relation:
\begin{eqnarray}
 J_\varphi(x) 
 =  - {\delta \Gamma[\varphi] \over \delta \varphi(x)}.
 \label{EAderivative}
\end{eqnarray}
The artificially introduced source $J$ has to vanish in
order to recover the original theory, i.e.
\begin{eqnarray}
 0
 =  - {\delta \Gamma[\varphi] \over \delta \varphi(x)} .
 \label{EAderivative0}
\end{eqnarray}
If the operator $O$ has been chosen so as to break the
symmetry of ${\cal L}$,
the nontrivial solution $\varphi \not= 0$ of this equation
gives the spontaneous symmetry breaking solution.
For the translation-invariant order parameter, we can
define the effective potential $V(\varphi)$ by
\begin{eqnarray}
 V(\varphi) := 
 - {\Gamma[\varphi(x) \equiv \varphi] \over \Omega},
 \quad
 \Omega := \int d^D x .
\end{eqnarray}
Note that the local order parameter is allowed to be a
composite operator, e.g.
\begin{eqnarray}
 O(x) = \bar \psi(x) \psi(x).
\end{eqnarray}
This case has been treated in sections 4 and 5.

\subsection{inversion formula}
\par
Suppose that the order parameter $\varphi$ is a function
of the source (i.e. probe) $J$ and is calculated in a
perturbative power series in $g$:
\begin{eqnarray}
 \varphi = f[J] = \sum_{n=0}^{\infty} g^n f_n[J].
 \label{series}
\end{eqnarray}
This is called the original series.  We know how to
calculate each term of the RHS of Eq.~(\ref{series}), since
this is nothing but the perturbation series.
\par
We need the inverted
function of Eq.~(\ref{series}), i.e.
$J$ is a function of $\varphi$:
\begin{eqnarray}
 J = f^{-1}[\varphi] .
 \label{inv}
\end{eqnarray}
We frequently encounter the situation in which it impossible
to obtain the exact form of the inverse function $f^{-1}$
explicitly.  However, we can get the above function
(\ref{inv}) by a perturbative power series of
$g$:
\begin{eqnarray}
 J = h[\varphi] = \sum_{n=0}^{\infty} g^n h_n[\varphi] ,
 \label{invseries}
\end{eqnarray}
which we call the inverted series.
The important point is that in Eq.~(\ref{invseries})
$\varphi$ is regarded as the order of unity.
Substituting Eq.~(\ref{invseries}) into Eq.~(\ref{series})
and expanding the RHS in a power series in $g$, we obtain
\begin{eqnarray}
 \varphi &=& f[h[\varphi]] 
 \nonumber\\
&=& \sum_{n=0}^{\infty} g^n f_n[
 \sum_{m=0}^{\infty} g^m h_m[\varphi]]
 \nonumber\\
&=& f_0[h_0[\varphi]+g h_1[\varphi] + g^2 h_2[\varphi]+...]
 \nonumber\\
&& + g f_1[h_0[\varphi]+g h_1[\varphi]+...]
 + g^2 f_2[h_0[\varphi]+g h_1[\varphi]+...]
 + ...
 \nonumber\\
&=&  f_0[h_0[\varphi]] 
 + g \{ f_0'[\varphi]h_1[\varphi]+ f_1[h_0[\varphi]] \}
 \nonumber\\
&& + g^2 \{ f_0'[h_0[\varphi]]h_2[\varphi]
+ {1 \over 2} f_0''[[h_0[\varphi]]h_1[\varphi]
\nonumber\\
&& \quad
+ f_1'[h_0[\varphi]]h_1[\varphi] + f_2[h_0[\varphi]] \}
 + ...,
 \label{consistency}
\end{eqnarray}
where the prime denotes the differentiation with respect
to $J$.  On the RHS of Eq.~(\ref{consistency}), $\varphi$
is regarded as order unity so that $\varphi$ on the LHS of
Eq.~(\ref{consistency}) should be regarded as order unity.
Then we can determine $h_m[\varphi]$ by equating each
power of $g$ on both sides of Eq.~(\ref{consistency}) and
get the inversion formula:
\begin{eqnarray}
 h_0[\varphi] &=& f_0^{-1}[\varphi],
 \nonumber\\
 h_1[\varphi] &=& - \left[ {f_1[J] \over
f_0'[J]} \right] \Biggr|_{J=h_0[\varphi]},
 \nonumber\\
 h_2[\varphi] &=& - \left[ {{1 \over 2}f_0''[J]
h_1^2[\varphi]+f_1'[J]h_1[\varphi]+f_2[J]
\over f_0'[J]} \right] \Biggr|_{J=h_0[\varphi]}, ...
 \label{invformula}
\end{eqnarray}
\par
Once we know the inverted series 
Eq.~(\ref{invseries}) according to the inversion formula
Eq.~(\ref{invformula}), the effective action is obtained
from Eq.~(\ref{EAderivative}) as follows.
\begin{eqnarray}
 \Gamma[\varphi] = - \int^\varphi d\phi h[\phi]
=  - \sum_{n=0}^{\infty} g^n
\int^\varphi d\phi h_n[\phi] .
\end{eqnarray}

\subsection{nonlocal source}

The inversion method can be applied to the nonlocal order
parameter as well as the local one.
For example, in order to
calculate the nonlocal order parameter 
$\varphi(x,y) := \langle O(x,y) \rangle$, 
We can add the source action
\begin{eqnarray}
{\cal S}_J = \int d^D x \int d^D y J(x,y) O(x,y).
\end{eqnarray}
In this case, we have
\begin{eqnarray}
  \Gamma[\varphi] 
  := W[J] - \int d^D x \int d^D y J(x,y)  \varphi(x,y),
\end{eqnarray}
\begin{eqnarray}
   \varphi(x,y) \equiv {\delta W[J] \over \delta J(x,y)} ,
\end{eqnarray}
and
\begin{eqnarray}
     J(x,y) 
  = - {\delta \Gamma[\varphi] \over \delta \varphi(x,y)}.
\end{eqnarray}
If we choose the fermion propagator as a nonlocal order
parameter:
\begin{eqnarray}
 \varphi(x,y) := \langle \bar \psi(x) \psi(y) \rangle
 := S(x,y),
\end{eqnarray}
the lowest order inversion in the coupling constant gives
the Schwinger-Dyson equation in the ladder approximation. 
Actually, in QED, we get (see \cite{UKF90}) in Minkowski
spacetime:
\begin{eqnarray}
 J(p) = i S^{-1}(p) - i S_0^{-1}(p) - ie^2 
 \int {d^Dk \over (2\pi)^D} \gamma^\mu S(p+k)\gamma^\nu
 D^{(0)}_{\mu\nu}(k) ,
\end{eqnarray}
where $D^{(0)}_{\mu\nu}(k)$ is the bare photon propagator.
If we set $J=0$, this coincides with the ladder SD equation
for the fermion propagator.
Using the following identification: 
\begin{eqnarray}
  J(p)  = - {\delta \Gamma[S] \over \delta S(p)} ,
\end{eqnarray}
we get the CJT effective action
\begin{eqnarray}
 \Gamma[S] &=& - i \int {d^Dk \over (2\pi)^D}
 \tr[\ln S(p)  - S(p) S_0^{-1}(p)]
 \nonumber\\&&
+ {1 \over 2} ie^2 
 \int {d^Dk \over (2\pi)^D} \int {d^Dk \over (2\pi)^D} 
 \tr[\gamma^\mu S(p+k)\gamma^\nu S(p)] 
 D^{(0)}_{\mu\nu}(k) .
\end{eqnarray}
This is the case treated in sections 7 and 8.

\newpage
\centerline{\large {\bf Figure Captions}}

\begin{enumerate}
\item[Figure 1:]
The schematic phase diagram of the gauged Thirring model
in 3+1 dimensions at $m_0=0$.

\vskip 0.5cm
\item[Figure 2:]
The shape of the effective potential when $\tau=0$.
(a) Thirring$_4$ model, 
(b) QED$_4$, 
(c) gauged Thirring$_4$ model.

\vskip 0.5cm
\item[Figure 3:]
The critical line in the phase diagram $(\beta, \kappa)$ 
obtained from the gauge-invariant effective
potential.

\vskip 0.5cm
\item[Figure 4:]
The graphical representation for 
(a) the CJT effective action, and
(b) the SD equation obtained from the CJT effective action.

\vskip 0.5cm
\item[Figure 5:]
Plot of the nonlocal gauge-fixing function $\eta(k^2)$.

\vskip 0.5cm
\item[Figure 6:]
The critical line in the phase diagram $(\beta_e,1/g)$
obtained from the Schwinger-Dyson equation.

\vskip 0.5cm
\item[Figure 7:]
Equi-$\xi$ (correlation length) or equi-$m_d/\Lambda$
(fermion dynamical mass) line 
in the phase diagram $(\beta_e,1/g)$
obtained from the Schwinger-Dyson equation.
The line is parameterized from above to below as
$\xi^{-1}:=m_d/\Lambda =$
$0.0$, $0.05$, $0.10$, $0.15$, $0.20$, $0.25$, $0.30$,
$0.35$, $0.40$.

\vskip 0.5cm
\item[Figure 8:]
Definition of the renormalized coupling constant
$\alpha_R$.

\vskip 0.5cm
\item[Figure 9:]
The equi-$\alpha_R$ (renormalized coupling) line  
in the phase diagram $(\beta_e,1/g)$
obtained from the Schwinger-Dyson equation.
The line is parameterized from above to below as
$\alpha_R =$
$0.0$, $0.1$, $0.2$, $0.3$, $0.4$, $0.6$, $0.8$, $1.0$,
$1.5$, $2.0$, $2.5$, $3.0$.

\vskip 0.5cm
\item[Figure 10:]
The equi-$r_m$ (mass ratio) line 
in the phase diagram $(\beta_e,1/g)$
obtained from the Schwinger-Dyson equation.
The line is parameterized from above to below as
$r_m =$
$0.0$, $0.5$, $1$, $2$, $3$, $10$, $100$.

\end{enumerate}

\end{document}